\documentclass[runningheads]{llncs}

\usepackage[T1]{fontenc}
\usepackage{amsmath,amssymb,amsfonts}
\usepackage{mathtools}
\usepackage{stmaryrd}  % for \llbracket, \rrbracket
\usepackage{xspace}
\usepackage{xcolor}
\usepackage{listings}
\usepackage{hyperref}
\usepackage[capitalize]{cleveref}
\usepackage{comment}
\usepackage{longtable}
\usepackage{booktabs}

\usepackage{thmtools}
\usepackage{thm-restate}

\makeatletter

\let\c@theorem\relax

\let\c@definition\relax

\let\c@lemma\relax

\let\c@corollary\relax

\let\c@proposition\relax

\let\c@example\relax

\let\c@remark\relax
\makeatother

\declaretheorem[numberwithin=section, name=Definition]{definition}
\declaretheorem[sibling=definition, name=Example]{example}
\declaretheorem[sibling=definition, name=Theorem]{theorem}
\declaretheorem[sibling=definition, name=Proposition]{proposition}
\declaretheorem[sibling=definition, name=Lemma]{lemma}
\declaretheorem[sibling=definition, name=Corollary]{corollary}
\declaretheorem[sibling=definition, name=Remark]{remark}

\newcommand{\ia}{\textit{i}}
\newcommand{\ib}{\textit{ii}}
\newcommand{\ic}{\textit{iii}}

\newcounter{pc}

\newcommand{\calV}{\mathcal{V}}
\newcommand{\calT}{\mathcal{T}}
\newcommand{\calG}{\mathcal{G}}
\newcommand{\calC}{\mathcal{C}}

\lstdefinelanguage{glp}{
  morekeywords={},
  morecomment=[l]{\%},
  morestring=[b]",
  literate={:-}{{:\textrm{-}}}2 {?}{{?}}1,
}
\newcommand{\mypara}[1]{\vspace{1pt}\noindent\textbf{#1.}}
\newcommand{\temph}[1]{\textbf{#1}}
\newcommand{\remove}[1]{}

\newif\ifappendix
\appendixtrue

\newcommand{\arxivref}{the full paper~\cite{shapiro2026types}}

\usepackage{etoolbox}
\BeforeBeginEnvironment{verbatim}{\vspace{-0.6\baselineskip}}
\AfterEndEnvironment{verbatim}{\vspace{-0.5\baselineskip}}

\makeatletter
\renewcommand\section{\@startsection{section}{1}{\z@}%
  {-6\p@ \@plus -2\p@ \@minus -2\p@}%
  {3\p@ \@plus 1\p@}%
  {\normalfont\bfseries\large}}
\renewcommand\subsection{\@startsection{subsection}{2}{\z@}%
  {-4\p@ \@plus -1\p@}%
  {2\p@ \@plus 1\p@}%
  {\normalfont\bfseries}}
\renewcommand\subsubsection{\@startsection{subsubsection}{3}{\z@}%
  {-3\p@ \@plus -1\p@}%
  {1\p@ \@plus 1\p@}%
  {\normalfont\bfseries}}
\makeatother

\AtBeginDocument{%
  \setlength{\abovedisplayskip}{2pt plus 1pt minus 1pt}%
  \setlength{\belowdisplayskip}{2pt plus 1pt minus 1pt}%
  \setlength{\abovedisplayshortskip}{1pt plus 1pt}%
  \setlength{\belowdisplayshortskip}{1pt plus 1pt}%
}

\makeatletter
\def\thm@space@setup{%
  \thm@preskip=0pt plus 1pt minus 1pt
  \thm@postskip=\thm@preskip
  \interlinepenalty=0
  \clubpenalty=0
  \widowpenalty=0
}
\makeatother

\usepackage{enumitem}
\setlist{nosep, leftmargin=*, itemsep=0pt, topsep=0pt}

\title{Moded Types for Grassroots Logic Programs,\texorpdfstring{\\}{ }by AI, for AI\ifappendix\texorpdfstring{\\}{ }(Full Version)\fi}
\titlerunning{Moded Types for Grassroots Logic Programs}

\author{Ehud Shapiro}
\authorrunning{E.~Shapiro}
\institute{London School of Economics and Weizmann Institute of Science}

\begin{document}
\maketitle

\begin{abstract}
Grassroots Logic Programs (GLP) is a concurrent logic programming language in which logic variables are partitioned into paired readers and writers. An assignment is produced at most once via a writer and consumed at most once via its paired reader, and may contain additional readers and/or writers.  This enables the concise expression of rich multidirectional communication modalities.

``Logic Programs as Types for Logic Programs'' (LICS'91) defined types as regular sets of paths over the Herbrand atom semantics of a logic program.  Here, we develop a \emph{moded-atom semantics} that extends the standard Herbrand atom semantics in two ways: (\ia)~each atom subterm carries a \emph{mode}, recording whether it is consumed from or produced to the environment; and (\ib)~partial computations, including those that deadlock, fail, or never terminate, also contribute moded atoms to the semantics.  We define types to be regular sets of \emph{moded paths} over this semantics, give a syntactic definition of GLP well-typing, and prove that a well-typed program is sound: every output path in its well-typed moded-atom semantics conforms to its declared output type.

A type checker for GLP was implemented \emph{by} AI (Claude) in Dart, starting from the mathematical specification of Typed GLP (this paper), deriving from it an English+pseudocode spec (written by AI), and from the spec deriving Dart code (by AI). While GLP is naturally untyped, the motivation for typing it was \emph{for} AI: tasking AI to program complex communication modalities and hoping for the best turned out to be a tenuous strategy. The discipline we developed with Typed GLP is for the human designer and AI to jointly develop formal GLP type definitions and declarations, together with informal intent of the declared procedures, and only then let AI write the GLP code.  This way, subtle mode errors that hitherto required tedious joint human-AI   debugging are caught by the type checker, with the net result that following the initial human-AI agreement on types, AI can develop,  test, and debug Typed GLP programs with design-level oversight from the human rather than code-level intervention.
\end{abstract}

% Include sections
%==============================================================================
\section{Introduction}
\label{sec:introduction}
%==============================================================================

\mypara{Grassroots Logic Programs}
Grassroots Logic Programs (GLP) is a concurrent logic programming language in which logic variables are partitioned into paired \emph{readers} and \emph{writers}, recalling both linear logic~\cite{girard1987linear} and futures/promises~\cite{baker1977future,friedman1976impact}: an assignment is produced at most once via the sole occurrence of a writer (promise) and consumed at most once via the sole occurrence of its paired reader (future), and may contain additional readers and/or writers.  

Hence, GLP provides for the concise expression of multidirectional communication modalities, including bidirectional communication and the dynamic expansion and  reconfiguration of communication networks.  
GLP's design eschews unification in favour of simple term matching. Being relational and nondeterministic, GLP---like its predecessor concurrent logic languages~\cite{shapiro1989family}---can express nondeterministic asynchronous concurrent processes such as the fair merging of streams.
GLP is designed for the implementation of smartphone-based, serverless, grassroots platforms~\cite{shapiro2025glp}.  A digital platform is \emph{grassroots}~\cite{shapiro2023grassrootsBA,shapiro2025atomic} if it can have multiple instances that can (\ia)~operate independently of each other and of any global resource other than the network, and (\ib)~coalesce into ever larger instances, possibly resulting in a single global instance.  The SRSW restriction makes GLP suitable for grassroots distributed execution: each variable assignment has exactly one producer and one consumer, so assignments can be communicated between independent agents without consensus.

\mypara{Typing GLP}
``Logic Programs as Types for Logic Programs''~\cite{fruhwirth1991lics} defined types as regular sets of paths over the Herbrand atom semantics of a logic program.  In this paper, we develop a \emph{moded-atom semantics} that extends the standard Herbrand atom semantics in two ways: (\ia)~each atom subterm carries a \emph{mode}, recording whether it is consumed from or produced to the environment; and (\ib)~partial computations, including those that deadlock, fail, or never terminate, also contribute moded atoms to the semantics.  We define types to be regular sets of \emph{moded paths} over this semantics, give a syntactic definition of GLP well-typing, and prove that a well-typed program is sound: every output path in its well-typed moded-atom semantics conforms to its declared output type.

Types are specified using BNF rules with a complementation operator~$?$ that relates producer and consumer views: a type \verb|Stream| implicitly defines its dual \verb|Stream?|, with modes complemented throughout.  Procedure type declarations specify the mode of each argument, e.g., \verb|merge(Stream?,Stream?,Stream)| declares that \verb|merge| consumes two streams and produces one.

A GLP program is \emph{well-typed} if three conditions hold.  First, \emph{output conformance}: each clause's head and body must have paths accepted by the declared types.   Second, \emph{input coverage}: every input path permitted by the type must be accepted by some clause.  Third, \emph{variable-pair compatibility}: a writer~$X$ and its paired reader~$X?$ have dual types, relaxed in Section~\ref{sec:subtyping} to the type of~$X$ being a \emph{subtype} of the dual of the type of~$X?$~\cite{gay2005subtyping}.

The \emph{well-typed moded-atom semantics} of a program is the set of moded atoms its runs produce from \emph{well-typed initial goals}---goals whose arguments conform to the declared types; it makes no claim about runs from ill-typed goals.  The soundness of well-typing over this semantics is the \emph{covariant}, or subsumption, direction~\cite{pierce2002types,cardelli1985understanding}.

\mypara{Programming with AI}
AI is used here in two distinct disciplines: \emph{implementing} the GLP type checker, and \emph{programming} in GLP.

A type checker for Typed GLP was implemented in Dart~\cite{dart2024} \emph{by} AI (Claude) via a three-tier methodology: (\ia)~the mathematical specification of Typed GLP (this paper); (\ib)~an English+pseudocode spec, derived from (\ia) by AI; (\ic)~Dart code, derived from (\ib) by AI.  The implementation performs type AST and DFA construction, well-typed term and clause checking, subtyping, parameterised-type expansion with parameter-instantiation inference, and construction of moded heads from declarations; it is validated by positive and negative type-checking tests, SRSW-violation tests, and invalid-guard tests over a corpus of typed example programs. The methodology is iterative in both directions: gaps surfaced during code generation drove revisions to the specification and, in turn, to the definitions of this paper. 

This is the sense in which moded types are \emph{for} AI: they are a specification language for data types and communication modalities, providing a precise interface between the human designer and the AI programmer~\cite{fowler2025sdd,mundler2025type,blinn2024typed}.
The discipline we developed with Typed GLP is for the human designer and AI to jointly develop the formal type definitions and declarations together with the informal intent of the declared procedures, and only then let AI write the GLP code.  Mode errors---where a writer and a reader are confused---are a characteristic class of error in GLP; untyped, they surface only at run time, as silent suspensions, and require expensive joint human-AI debugging to locate.  Against a settled type system they are caught by the type checker at compile time, and AI corrects them from the diagnostic: for example, in a befriending program from the implementation corpus, a clause head was rejected at compile time---``Variable mode mismatch: reader requires ↓ (consume), got ↑ (produce); Path: (Decision?,\,0,\,output)''---and corrected with no human debugging.  Following the initial agreement on types, declarations, and intent, AI develops, type-checks, tests, and debugs the GLP code, with the human providing design-level oversight, including language-design decisions, rather than code-level intervention.  The implementation, with its type-checking test corpus, is publicly available.\footnote{\url{https://github.com/EShapiro2/GLP}}

\mypara{Paper Structure}
Section~\ref{sec:glp} presents GLP, its operational semantics, and the moded-atom semantics underlying the type system.  Section~\ref{sec:typed-glp} introduces moded types and procedure declarations.  Section~\ref{section:well-typing} defines GLP well-typing and its subtyping relaxation.  Section~\ref{sec:glp-semantics-well-typing} proves the soundness of well-typing.  Section~\ref{sec:parameterized-types} adds parameterised types by expansion.  All proofs appear in \ifappendix Appendix~\ref{app:proofs}\else the full paper~\cite{shapiro2026types}\fi.

\ifappendix
%==============================================================================
\section{Logic Programs}
\label{sec:lp}
%==============================================================================

We recall standard Logic Programs (LP) notions of syntax and semantics via goal reduction.  
%We use $\subset$ to denote the strict subset relation, $\subseteq$ when equality is also possible, and $a\ne b \in S$ as a shorthand for $a\ne b\wedge a\in S \wedge b\in S$.

%------------------------------------------------------------------------------
\subsection{Syntax}
\label{sec:lp-syntax}
%------------------------------------------------------------------------------

We employ the standard LP notions of variables, constants, terms, clauses, procedures, and programs.
%let $\calV$ denote the set of all variables,  $\calT$ the set of all terms, and $\calG(P)$ the set of all goals over a program $P$.

\begin{definition}[Logic Programs Syntax]
\label{def:lp-syntax}
Let $\calV$ denote the set of \temph{variables} (identifiers beginning with uppercase).  A \temph{term} is a variable, a constant (numbers, strings, or the empty list \verb|[]|), or a compound term $f(T_1,\ldots,T_n)$ with functor $f$ and subterms $T_i$.  Let $\calT$ denote the set of all terms.  We use standard list notation: \verb=[X|Xs]= for a list cell, \verb|[X1,...,Xn]| for finite lists.  A term is \temph{ground} if it contains no variables.

A \temph{unit goal} is a compound term or a string, also called an \temph{atom}.  A \temph{goal} is a multiset of unit goals; the empty goal is written \verb|true|.  A \temph{clause} $A$~\verb|:-|~$B$ has head $A$ (a unit goal) and body $B$ (a goal); a \temph{unit clause} has empty body.  A \temph{logic program} is a finite set of clauses; clauses for the same predicate form a \temph{procedure}.  Let $\calG(P)$ denote the set of goals over program $P$.
\end{definition}

\ifappendix
\begin{example}[Append]
The standard logic program for list concatenation is the following procedure:
\begin{verbatim}
append([X|Xs], Ys, [X|Zs]) :- append(Xs, Ys, Zs).
append([], Ys, Ys).
\end{verbatim}
Logically, a clause $A$ \verb|:-| $B$ is a universally-quantified implication in which $B$ implies $A$, and a program is a conjunction of its clauses.
By convention, we use plural variable names like \verb|Xs| to denote a list of \verb|X|'s.
\end{example}
\fi

%------------------------------------------------------------------------------
\subsection{Operational Semantics}
\label{sec:lp-operational}
%------------------------------------------------------------------------------
A \emph{substitution} $\sigma$ is an idempotent function $\sigma: \calV \to \calT$. By convention, $\sigma(x)=x\sigma$. Let $\Sigma$ denote the set of all substitutions. We assume the standard notions of instance, ground, renaming, renaming apart, unifier, and most-general unifier (mgu). We assume a fixed renaming-apart function, so that the result of renaming $T'$ apart from $T$ is well defined.

\begin{definition}[LP Goal/Clause Reduction]
\label{def:lp-reduction}
Given an LP unit goal $A$ and clause $C$, with $H$ \verb|:-| $B$ being the result of renaming $C$ apart from $A$, the \temph{LP reduction} of $A$ with $C$ \temph{succeeds with} $(B,\sigma)$ if $A$ and $H$ have an mgu $\sigma$.
\end{definition}
We define the operational semantics of LP and of GLP via transition systems; in both, set relations and operations refer to multisets.

\begin{definition}[Transition System~\cite{shapiro2025glp}]
\label{def:transition-system}
A \temph{transition system} is a tuple $TS = (C, c_0, T)$ where:
\begin{itemize}
    \item $C$ is an arbitrary set of \temph{configurations}
    \item $c_0 \in C$ is a designated \temph{initial configuration}
    \item $T \subseteq C \times C$ is a \temph{transition relation}. A transition $(c,c') \in T$ is also written as $c \rightarrow c' \in T$.
\end{itemize}
A transition $c \rightarrow c' \in T$ is \temph{enabled} from configuration $c$. A configuration $c$ is \temph{terminal} if no transitions are enabled from $c$. A \temph{computation} is a (finite or infinite) sequence of configurations where for each two consecutive configurations $(c,c')$ in the sequence, $c \rightarrow c' \in T$. We write $c \xrightarrow{*} c'$ to denote the existence of a computation (empty if $c = c'$) from $c$ to $c'$. A \temph{run} is a computation starting from $c_0$, which is \temph{complete} if it is infinite or ends in a terminal configuration. The \temph{outcome} of a complete run is determined by a domain-specific function from complete runs to an outcome space.
\end{definition}

\begin{definition}[Logic Programs Transition System]
\label{def:lp-ts}
A transition system $LP(P) = (C, c_0, T)$ is a \temph{Logic Programs transition system} for a logic program $P$ and initial goal $G_0 \in \mathcal{G}(P)$, if $C=\mathcal{G}(P)\times \Sigma$, $c_0=(G_0,\emptyset)$, and $T$ is the set of all transitions $(G,\sigma) \rightarrow (G',\sigma') \in (\mathcal{G}(P)\times \Sigma)^2$ such that for some unit goal $A \in G$ and clause $C \in P$ the LP reduction of $A$ with $C$ succeeds with $(B,\hat\sigma)$,  $G' = (G \setminus \{A\} \cup B)\hat\sigma$, and $\sigma'=\sigma\circ\hat\sigma$.
\end{definition}

\begin{comment}
As a tribute to resolution theorem proving~\cite{robinson1965machine}---the intellectual ancestor of logic programming---a configuration of $LP$ is also referred to as a \emph{resolvent}.
Logic Programs have two forms of nondeterminism: the choice of $A\in G$, called \emph{and-nondeterminism}, and the choice of $C\in P$, called \emph{or-nondeterminism}. Thus, as an abstract model of computation, LP are closely-related to \emph{Alternating Turing Machines}, a generalization of Nondeterministic Turing Machines~\cite{shapiro1984alternation}.
\end{comment}

The following notion of a proper run ensures that variable names are not re-used.
\begin{definition}[Proper and Successful Run, Outcome~\cite{shapiro2025glp}]
\label{def:proper-run}
A run $\rho: (G_0,\sigma_0) \rightarrow  \cdots \rightarrow (G_n, \sigma_n)$ of $LP(P)$ is \temph{proper} if for any $1\le i< n$, a variable that occurs in $G_{i+1}$ but not in $G_i$ also does not occur in any $G_j$, $j<i$. If proper, the \temph{outcome} of $\rho$ is $(G_0$ \verb|:-| $G_n)\sigma_n$. Such a run is \temph{successful} if $G_n=\emptyset$.
\end{definition}

\ifappendix
The following proposition justifies calling a proper LP run a \emph{derivation}, and a complete proper run ending in the empty goal a \emph{successful derivation}.

\begin{restatable}[LP Computation is Deduction]{proposition}{propLPDeduction}
\label{prop:lp-deduction}
The outcome $(G_0 \mathrel{\mbox{\texttt{:-}}} G_n)\sigma$ of a proper run $\rho: (G_0,\sigma_0) \rightarrow  \cdots \rightarrow (G_n, \sigma_n)$ of $LP(P)$  is a logical consequence of $P$.
\end{restatable}

Note that the LP transition system does not require the initial goal $G_0$ to be a unit goal.
\subsection{Denotational Semantics}

The $LP(P)$ transition system allows defining several denotational semantic notions for a program $P$: 
\begin{enumerate}
    \item The \emph{clause semantics} of $P$ is the set of all outcomes of all proper runs of $LP(P)$ with an initial most-general unit goal (arguments are distinct variables).
    It is closely related to the fully-abstract compositional semantics of LP~\cite{gaifman1989fully}.
    \item The \emph{atom semantics} of $P$ is the set of all outcomes of all successful derivations of $LP(P)$ with an initial most-general unit goal.
    \item The \emph{ground atom semantics} of $P$ is the standard model-theoretic semantics of logic programs.  It is the set of ground instances of the atom semantics of $P$ over the Herbrand universe of $P$.
\end{enumerate}
\fi

\subsection{Types}

We associate with each logic term $T$ a \emph{term tree}:  a labelled tree with vertices labelled with the functor/arity for compound terms and constants/variables for leaves in $T$, and edges labelled with the argument index of the subterm.

The \emph{paths} of a term $T$, $paths(T)$, is the set of labelled paths in its term tree.  For example, in \verb=append([X|Xs],Ys,[X|Zs])=, the first argument contributes the path \verb|append --1--> "."/2 --1--> X|.

In~\cite{fruhwirth1991lics}, a type for LP is defined as a regular set of paths, and a program is well-typed if the type includes all paths of all atoms in its ground-atom semantics.  Types are specified using well-founded BNF rules and procedure type declarations, for example:

\begin{verbatim}
List ::= [] ; [_|List].

procedure append(List,List,List).
\end{verbatim}
with \verb|_| denoting any path in any ground term in the Herbrand universe of the program.  Note that LP well-typing is \emph{covariant}: it guarantees that all outputs are within the type, but does not constrain inputs. Thus \verb|append([],c,c)| is in the ground-atom semantics, and \verb|append| is well-typed by \verb|append(_,_,_)|.
So, providing useful type declarations is up to the programmer or program designer:  the more informative and tight the type declaration, the more useful is the statement that the program is well-typed.  \ifappendix Many extensions for this notion of LP types are possible and needed, in particular polymorphic types, discussed in~\cite{yardeni1991polymorphically}.\fi
\fi
%==============================================================================
\section{Grassroots Logic Programs}
\label{sec:glp}
%==============================================================================

Grassroots Logic Programs (GLP) extend LP by (\ia) adding a paired \emph{reader} $X?$ to every ``ordinary'' logic variable $X$, now called a \emph{writer}; (\ib) restricting variables in goals and clauses to have at most a single occurrence (SO); and (\ic) requiring that a variable occurs in a clause iff its paired variable also occurs in it (single-reader single-writer, SRSW). The result eschews unification in favour of simple term matching, is linear-logic-like~\cite{girard1987linear}, and is futures/promises-like~\cite{baker1977future,friedman1976impact}: each assignment $X := T$ is produced at most once via the sole occurrence of a writer (promise) $X$, and consumed at most once via the sole occurrence of its paired reader (future) $X?$.

%------------------------------------------------------------------------------
\subsection{Syntax}
\label{sec:glp-syntax}
%------------------------------------------------------------------------------

\begin{definition}[GLP Variables]
\label{def:glp-variables}
\ifappendix Recall that\else Let\fi\ $\calV$ \ifappendix is\else be\fi\ the set of LP variables, henceforth called \temph{writers}. Define $\calV? = \{X? \mid X \in \calV\}$, called \temph{readers}. The set of all GLP variables is $\hat\calV = \calV \cup \calV?$. A writer $X$ and its reader $X?$ form a \temph{variable pair}.
\end{definition}

GLP terms, unit goals, goals, and clauses are as in LP\ifappendix\ (Definition~\ref{def:lp-syntax})\fi, but defined over the variables in $\hat\calV$.

\begin{definition}[Single-Occurrence (SO) Invariant]
\label{def:so-invariant}
A term, goal, or clause satisfies the \temph{single-occurrence (SO) invariant} if every variable occurs in it at most once.
\end{definition}

\begin{definition}[GLP Program]
\label{def:glp-program}
A clause $C$ satisfies the \temph{single-reader/single-writer (SRSW) restriction} if it satisfies SO and a variable occurs in $C$ iff its paired variable also occurs in $C$.
A \temph{GLP program} is a finite sequence of clauses satisfying SRSW; clauses for the same predicate form a \temph{procedure}.
Let $\hat\calG(P)$ denote the set of goals over $\hat\calV$ and the vocabulary of $P$ that satisfy SO.
\end{definition}

\ifappendix As we shall see (Proposition~\ref{prop:so-preservation})\else As shown in \arxivref\fi, reducing a goal satisfying SO with a clause satisfying SRSW results in a goal satisfying SO.  The purpose of the SRSW restriction is to prevent multiple writer occurrences racing to assign a variable.

\begin{example}[Fair Merge]
\label{ex:merge}
Consider the standard concurrent logic program for fairly merging two streams, written in GLP:
\begin{verbatim}
merge([X|Xs], Ys, [X?|Zs?]) :- merge(Ys?, Xs?, Zs).
merge(Xs, [Y|Ys], [Y?|Zs?]) :- merge(Xs?, Ys?, Zs).
merge(Xs, [], Xs?).
merge([], Ys, Ys?).
\end{verbatim}
and the goal \verb=merge([1,2,3|Xs?],[a,b|Ys?],Zs)=. Both the goal and each clause satisfy SO, and each clause satisfies SRSW. The first two clauses swap inputs in recursive calls, ensuring fairness when both streams are available.
\end{example}

%------------------------------------------------------------------------------
\subsection{Operational Semantics}
\label{sec:glp-operational}
%------------------------------------------------------------------------------

As is standard for concurrent programming languages~\cite{plotkin1981structural}, we provide GLP with operational semantics using a nondeterministic transition system, termed \emph{cGLP} for concurrent GLP; the paper defining GLP~\cite{shapiro2025glp} also provides it with multiagent operational semantics.  The basic Reduce transition of cGLP (Definition~\ref{def:cglp-ts})  differs from LP in (\ia) using a writer mgu instead of a regular mgu---the fundamental use of GLP readers for synchronisation---and (\ib) choosing the first applicable clause instead of any clause to enable the expression of fair concurrent programs such as fair merge.   Communicate propagates an assignment from a writer to its paired reader. 

\begin{definition}[Writers Substitution, Assignment, Readers Substitution and Counterpart]
\label{def:writers-assignment}
A GLP \temph{writer assignment} is a term of the form $X := T$,  $X\in\calV$,  $T\notin\calV$,
satisfying SO. 
Similarly, a GLP \temph{reader assignment} is a term of the form $X? := T$,  $X?\in\calV?$,  $T\notin\calV$,
satisfying SO. 
A \temph{writers (readers) substitution} $\sigma$ is the substitution implied by a set of writer (reader) assignments that jointly satisfy SO. Given a writer assignment $X := T$, its \temph{readers counterpart} is $X? := T$, and given a writers substitution $\sigma$, its \temph{readers counterpart} $\sigma?$ is the readers substitution defined by $X?\sigma? = X\sigma$. Dually, given a reader assignment $X? := T$, its \temph{writers counterpart} is $X := T$, and given a readers substitution $\tau$, its \temph{writers counterpart} $\tau!$ is the writers substitution defined by $X\tau! = X?\tau$. The \temph{pair completion} of a readers substitution $\sigma$ is $\sigma^\star = \sigma \cup \sigma!$, applied to a fixed point. 
\end{definition}

\begin{definition}[GLP Renaming, Renaming Apart]
\label{def:glp-renaming}
A \temph{GLP renaming} is an injective substitution $\rho: \hat\calV \to \hat\calV$ such that for each $X \in \calV$: $X\rho \in \calV$ and $X?\rho = (X\rho)?$. 
Two GLP terms \temph{have a variable in common} if for some writer $X \in \calV$, either $X$ or $X?$ occurs in both. A GLP renaming $\rho$ \temph{renames $T'$ apart from} $T$ if $T'\rho$ and $T$  have no variable in common.
\end{definition}

\begin{definition}[Writer MGU~\cite{shapiro2025glp}]
\label{def:writer-mgu}
Given two GLP unit goals $A$ and $H$, a \temph{writer mgu} is a writers substitution $\sigma$ such that $A\sigma = H\sigma$ and $\sigma$ is most general among such substitutions.
\end{definition}

\begin{definition}[GLP Goal/Clause Reduction]\label{definition:GLP-goal-clause-reduction}
Given GLP unit goal $A$ and clause $C$,  with $H \verb|:-| B$ being the result of the GLP renaming of $C$ apart from $A$, the \temph{GLP reduction} of $A$ with $C$ \temph{succeeds with result} $(B,\sigma)$ if $A$ and $H$ have a writer mgu $\sigma$.
\end{definition}

\begin{definition}[cGLP Transition System~\cite{shapiro2025glp}]
\label{def:cglp-ts}
Given a GLP program $P$, an \temph{asynchronous resolvent} over $P$ is a pair $(G, \sigma)$ where $G \in \hat\calG(P)$ and $\sigma$ is a readers substitution.

A transition system $cGLP = (\calC, c_0, \calT)$ is a \temph{cGLP transition system} over $P$ and initial goal $G_0$ satisfying SO if:
\begin{enumerate}
    \item $\calC$ is the set of all asynchronous resolvents over $P$
    \item $c_0 = (G_0, \emptyset)$
    \item $\calT$ is the set of all transitions $(G, \sigma) \rightarrow (G', \sigma')$ satisfying either:
    \begin{enumerate}
        \item \textbf{Reduce:} there exists unit goal $A \in G$ such that $C \in P$ is the first clause for which the GLP reduction of $A$ with $C$ succeeds with result $(B, \hat\sigma)$, $G' = (G \setminus \{A\} \cup B)\hat\sigma$, and $\sigma' = \sigma \circ \hat\sigma?$
        \item \textbf{Communicate:} $\{X? := T\} \in \sigma$, $X?\in G$,
        $G' = G\{X? := T\}$, and $\sigma' = \sigma$
    \end{enumerate}
\end{enumerate}
\end{definition}

\begin{definition}[Run, Outcome]
\label{def:cglp-run-outcome}
A \temph{run} of $cGLP$ is a computation $(G_0,\sigma_0)\rightarrow\cdots\rightarrow(G_n,\sigma_n)$ starting from $c_0$, \temph{complete} if it is infinite or ends in a \temph{terminal} configuration, one from which no transition is enabled.  A run is \temph{proper} if for any $1\le i<n$, a variable that occurs in $G_{i+1}$ but not in $G_i$ does not occur in any $G_j$, $j<i$; the \temph{outcome} of a proper run is $(G_0$ \verb|:-| $G_n)\sigma_n^\star$, and the run is \temph{successful} if $G_n=\emptyset$.
\end{definition}

\ifappendix
\mypara{Monotonicity}
Key differences between LP and cGLP relate to monotonicity. In LP, if a goal cannot be reduced, it will never be reduced. In cGLP, a goal that cannot be reduced now may be reduced in the future: if $A$ and $H$ have an mgu that writes on a reader $X? \in A$, and therefore have no writer mgu at present,  another goal that has $X$ may reduce, assigning $X$, and later $X?$, to a value that will allow $A$ and $H$ to have a writer mgu.  Conversely, in LP, if a goal $A$ can be reduced now with some clause $H:-B$, with a regular mgu of $A$ and $H$, it may not be reducible in the future due to variables that $A$ shares with other goals being assigned values by other goal reductions, preventing unification between the instantiated $A$ and $H$.  In cGLP, if a goal $A$ can be reduced now (with a writers mgu), it can always be reduced in the future, as the SO invariant ensures that no other goal can assign any writer in $A$.

In an implementation, if a cGLP goal $A$ cannot be reduced now, but there is a readers substitution $\sigma$ such that $A\sigma$ can be reduced, such readers are identified, the goal $A$ \emph{suspends} on these readers, and is rescheduled for another reduction attempt once any of them is assigned.

Despite these differences, cGLP has the same notion of logical consequence as LP.  Let $/?$ be an operator that replaces every reader  by its paired writer.

\begin{restatable}[GLP Computation is Deduction]{proposition}{propGLPDeduction}
\label{prop:glp-deduction}
Let  $(G_0 \mathrel{\mbox{\texttt{:-}}} G_n)\sigma_n^\star$ be the outcome of a proper  run $\rho: (G_0,\sigma_0) \rightarrow  \cdots \rightarrow (G_n, \sigma_n)$ of $cGLP(P)$.
Then $(G_0 \mathrel{\mbox{\texttt{:-}}} G_n)\sigma_n^\star/?$ is a logical consequence of $P/?$.
\end{restatable}

\fi

GLP runs satisfy two safety properties.

\begin{restatable}[SO Preservation]{proposition}{propSOPreservation}
\label{prop:so-preservation}
If the initial goal $G_0$ satisfies SO, then every goal in a proper cGLP run satisfies SO.
\end{restatable}

\begin{restatable}[Monotonicity]{proposition}{propMonotonicity}
\label{prop:glp-monotonicity}
In any proper cGLP run, if unit goal $A$ can reduce with clause $C$ at step $i$, then either an instance of $A$ has been reduced by step $j > i$, or an instance of $A$ can still reduce with $C$ at step $j$.
\end{restatable}

\subsection{Term Matching Eschews Unification}

If two terms $T_1$ and $T_2$ that jointly satisfy SO are unifiable with an mgu $\sigma$, then $\sigma$ maps any variable in $T_1$ to a subterm of $T_2$ and vice versa.  Hence, the SO invariant of GLP allows eschewing unification in favour of \emph{term matching}  that performs joint term-tree traversal and collects variable assignments along the way, as follows.

\begin{definition}[Term Matching]
\label{def:term-matching}
Given a goal $T_1$ and a renamed-apart clause head $T_2$ that jointly satisfy SO, their \temph{term matching} proceeds via the joint traversal of the term-trees of $T_1$ and $T_2$, consulting the following table at each pair of joint vertices ($T_1$ ranging over the goal's subterms, $T_2$ over the head's), where $X_1, X_2$ denote writers, $X_1?, X_2?$ denote readers, and $f/n$ denotes a non-variable term, a constant when $n=0$ and a compound term when $n>0$:
\begin{center}
\begin{tabular}{l|lll}
$T_1 \backslash T_2$ & Writer $X_2$ & Reader $X_2?$ & Term $f_2/n_2$ \\
\hline
Writer $X_1$ & fail & $X_1 := X_2?$ & $X_1 := T_2$ \\
Reader $X_1?$ & $X_2 := X_1?$ & fail & suspend on $X_1?$\\
Term $f_1/n_1$ & $X_2 := T_1$ &  fail & fail if $f_1\ne f_2$ or $n_1\ne n_2$\\
\end{tabular}
\end{center}
The writer mgu is the union of all writer assignments if no \emph{fail} was encountered and the suspension set is empty.
\end{definition}

%------------------------------------------------------------------------------
\subsection{Moded-Atom Semantics}
\label{sec:moded-atom-semantics}
%------------------------------------------------------------------------------

The basic semantic object that captures the behaviour of a GLP program is the moded term, which specifies whether each subterm was produced or consumed during the computation.

\begin{definition}[Moded Term, Dual]
\label{def:moded-term}
Given a GLP term $T$, a \temph{moded term} $T'$ corresponding to $T$ is the result of adding one of two \temph{mode annotations}, \temph{consume} \verb|↓| or \temph{produce} \verb|↑|, to $T$ and to every non-variable subterm of $T$. Given a moded term $T$, its \temph{dual}  $T?$ is obtained by \emph{complementation}: complementing every mode annotation and replacing every variable by its paired variable.
\end{definition}

By convention, a writer carries the produce mode \verb|↑| and its paired reader the consume mode \verb|↓|.

For example, consider the term:
\begin{verbatim}
merge([3|Xs?],Ys?,[3|Zs])
\end{verbatim}
and a corresponding moded term:
\begin{verbatim}
↓merge(↓[↓3|Xs?],Ys?,↑[↑3|Zs])
\end{verbatim}
Its dual is:
\begin{verbatim}
↑merge(↑[↑3|Xs],Ys,↓[↓3|Zs?])
\end{verbatim}

Substitution extends to moded terms: for a moded term $T$ and substitution $\sigma$, the moded term $T\sigma$ is obtained by replacing each variable $X$ or $X? \in T \cap \mathrm{dom}\,\sigma$ with the term $X\sigma$ or $X?\sigma$, which inherits the mode of its position.

\begin{definition}[Moded-atoms outcome and moded resolvent of a run]
\label{def:moded-outcome}
Given a proper cGLP run of a GLP program $P=(Cs,D)$ with outcome $(G_0 \mathrel{\mbox{\texttt{:-}}} G)\sigma^\star$ (Definition~\ref{def:cglp-run-outcome}), let $G_0'$ be $G_0\sigma^\star$ moded per $D$: each consumed position carries mode~$\downarrow$ and each produced position mode~$\uparrow$. The \temph{moded-atoms outcome} is the set of unit goals of $G_0'$; the \temph{moded resolvent} $G'$ is $G\sigma^\star$ moded the same way.
\end{definition}

\begin{definition}[Moded-Atom Semantics]
\label{def:moded-atom-semantics}
The \temph{moded-atom semantics} of a GLP program $P$ is the set of all moded-atoms outcomes of every finite proper run of $P$ with an SO initial goal.
\end{definition}

%------------------------------------------------------------------------------
\ifappendix
\subsection{Guards}
\label{sec:guards}
%------------------------------------------------------------------------------

GLP clauses may include \emph{guards}---tests that determine clause applicability.

\begin{definition}[Guarded Clause]
\label{def:guarded-clause}
A \temph{guarded clause} has the form $H$ \verb|:-| $G$ \verb"|" $B$, where $H$ is the head, $G$ is a conjunction of guard predicates, and $B$ is the body.  The guard separator ``\verb"|"'' distinguishes guards from the body, and is interpreted logically as a conjunction.  Guard arguments are readers paired to head writers.
\end{definition}

Guards have three-valued semantics.  Each guard predicate defines its \emph{success} condition.  A guard \emph{suspends} if it does not succeed but some instance of it under a readers substitution would succeed.  A guard \emph{fails} if no such instance exists.  A guard conjunction succeeds if all members succeed; it suspends if any member suspends and none fail; it fails if any member fails.

Definition~\ref{definition:GLP-goal-clause-reduction} of a GLP goal/clause reduction is augmented to succeed if the guard also succeeds.

\begin{remark}[Guards and SRSW]
\label{rem:guards-srsw}
Guard occurrences count toward SRSW satisfaction: if $X?$ occurs in a guard, its paired writer $X$ must occur in the head and $X?$ may additionally occur once in the body.

Furthermore, if the success of a guard implies that $X?$ is bound to a ground term, then both $X$ and $X?$ may occur multiple times in the clause.  Groundness-implying guards include \verb|ground|, \verb|integer|, \verb|number|, \verb|string|, \verb|constant|, arithmetic comparisons (\verb|<|, \verb|>|, \verb|=<|, \verb|>=|, \verb|=:=|, \verb|=\=|), and ground equality (\verb|=?=|).  Guards such as \verb|known| and \verb|compound| do not imply groundness: \verb|known(X?)| succeeds when $X?$ is bound to \verb|f(Y?)| with $Y?$ unbound.
\end{remark}

Guard predicate type signatures appear in the companion paper~\cite{shapiro2025glp}.
\else
The full language also includes guards, the type-handling of which is specified in \arxivref.
\fi
\section{Typed GLP}
\label{sec:typed-glp}
While GLP is untyped, we introduce Typed GLP; well-typing follows in Section~\ref{section:well-typing}.

\subsection{Type Syntax}
\label{sec:type-syntax}

\subsubsection{Type Declarations}
\label{sec:type-declarations}
LP types are concerned with the values logic variables may take.  A GLP type also indicates whether a term, or a subterm, is produced or consumed.  By convention, we define types from the perspective of their producer, which implicitly defines its dual---the same type but from the perspective of its consumer.

GLP types are specified using BNF rules with the complementation operator~$?$. For example, the stream producer type:
\begin{verbatim}
Stream ::= [] ; [_|Stream].
\end{verbatim}
implicitly defines the dual stream consumer type \verb|Stream?|, with the primitive types \verb|_| and \verb|_?| as wildcards accepting any produced or consumed subtree, respectively.

A type that is not complemented (for example \verb|Stream|) is an \emph{output type}, and its dual (for example \verb|Stream?|) is an \emph{input type}. A type that contains both produced and consumed positions is an \emph{interactive type}.  An alternative in a type definition may also be a type name~$S$, providing type union for types with disjoint top-level functors.  This disjointness is required so that a term's top functor selects its alternative deterministically.  Two alternatives with the same functor are not allowed; a stream of integers interspersed with \verb|switch| markers cannot be written as
\begin{verbatim}
Items ::= [] ; [Integer|Items] ; [switch|Items].
\end{verbatim}
whose last two alternatives share \verb|cons|.  Factoring the element choice into its own type removes the clash:
\begin{verbatim}
Items ::= [] ; [Item|Items].
Item  ::= Integer ; switch.
\end{verbatim}

The primitive types are \verb|Integer|, \verb|Real|, \verb|String|, and \verb|Module|.  \verb|Number|, \verb|Constant|, and \verb|Exp| (arithmetic expressions) are defined in the root \verb|self.glp| (Appendix~\ref{sec:root-self}): \verb|Number ::= Integer ; Real| and \verb|Constant ::= Number ; String ; Module|.

The formal specification of the type automaton induced by a typed GLP program appears in Section~\ref{sec:type-automaton}.

\subsubsection{Procedure Declarations}
\label{sec:procedure-declarations}
Typed GLP includes a type declaration for each procedure.  For example, the \verb|merge| program could include the procedure declaration:
\begin{verbatim}
procedure merge(Stream?,Stream?,Stream).
\end{verbatim}
which is a syntactic shorthand for defining a type named \verb|Procedure| using the standard BNF syntax:
\begin{verbatim}
Procedure ::= merge(Stream?,Stream?,Stream).
\end{verbatim}
Beyond syntactic convenience, the procedure declaration specifies the mode of each argument, essential for the well-typing conditions of Section~\ref{section:well-typing}.

\ifappendix
To ease the specification of types, we include parameterised types, a limited form of type polymorphism. A parameterised type declaration of \verb|merge| could look like:
\begin{verbatim}
Stream(X) ::= [] ; [X|Stream(X)].

procedure merge(Stream(X)?,Stream(X)?,Stream(X)).
\end{verbatim}

\mypara{Declaration placement}
Type declarations may appear anywhere in a program before they are used.
A procedure declaration must appear immediately before its defining clauses, which must be contiguous.
For example:
\begin{verbatim}
Stream ::= [] ; [_|Stream].

procedure merge(Stream?,Stream?,Stream).
merge([X|Xs], Ys, [X?|Zs?]) :- merge(Ys?, Xs?, Zs).
merge(Xs, [Y|Ys], [Y?|Zs?]) :- merge(Xs?, Ys?, Zs).
merge(Xs, [], Xs?).
merge([], Ys, Ys?).

procedure copy(Stream?, Stream).
copy([X|Xs], [X?|Ys?]) :- copy(Xs?, Ys).
copy([], []).
\end{verbatim}
\fi

\subsection{Program Syntax}
\label{sec:program-syntax}

\ifappendix\subsubsection{SRSW Relaxations}\else\mypara{SRSW Relaxations}\fi
\label{sec:srsw-relaxations}

The SRSW restriction (Definition~\ref{def:glp-program}) may be relaxed under certain conditions\ifappendix.  Section~\ref{sec:guards} described guard-based relaxations; here we describe additional relaxations available in Typed GLP.\else\ specified in \arxivref.\fi

\ifappendix
\mypara{Anonymous variables}  An \emph{anonymous variable} is any variable whose name begins with \verb|_| (e.g., \verb|_|, \verb|_In|, \verb|_Out|).  Anonymous variables may appear anywhere a writer variable may appear.  Each occurrence denotes a fresh writer with no paired reader, providing a controlled exception to the SRSW restriction.  Values assigned to anonymous variables are discarded.  For example:
\begin{verbatim}
second([_, X | _], X?).
foo(X) :- bar(_Result, X?).
\end{verbatim}
In the first clause, \verb|_| discards the head and tail of the input list.  In the second, \verb|_Result| discards the first output of \verb|bar|.

The symbols \verb|_| and \verb|_?| in type definitions are primitive type symbols, not variables.
\fi

\ifappendix
\mypara{Type checking of guards}  For the purpose of type checking, guards are treated as conjunction with the body: \verb=H :- G | B= is type-checked as \verb=H :- G, B=.
\fi

\subsubsection{Typed GLP Programs}
\label{sec:typed-glp-programs}

\begin{definition}[Typed GLP Program]
\label{def:typed-glp-program}
A \temph{typed GLP program} $P=(Cs,D)$ consists of GLP clauses $Cs$ and type declarations $D$ such that:
\begin{enumerate}
    \item Every procedure in $Cs$ has exactly one type declaration in $D$.
    \item Every procedure declared in $D$ is defined by at least one clause in $Cs$.
    \item Every type referenced in $D$ is a primitive type or is defined by a type rule in $D$.
\end{enumerate}
\end{definition}

\ifappendix
\subsection{Typed GLP Programming Examples}
\label{sec:typed-glp-examples}

\mypara{Stream merging} Combining the elements above provides a typed GLP \verb|merge| program:
\begin{verbatim}
Stream ::= [] ; [_|Stream].

procedure merge(Stream?,Stream?,Stream).
merge([X|Xs], Ys, [X?|Zs?]) :- merge(Ys?, Xs?, Zs).
merge(Xs, [Y|Ys], [Y?|Zs?]) :- merge(Xs?, Ys?, Zs).
merge(Xs, [], Xs?).
merge([], Ys, Ys?).
\end{verbatim}

This \verb|merge| program is well-typed (Section~\ref{section:well-typing}).
It is also well-typed with weaker type declarations such as \verb|procedure merge(_?,_?,_)|, but is not well-typed with declarations that violate mode requirements, such as \verb|procedure merge(_,_,_)|.

So, as in LP, providing useful type declarations is up to the programmer. 

\mypara{Monitors with hollow messages} Next, we show an interactive type with mode complementation.  Consider a monitor process, maintaining a local state and serving requests to update the state and read it.  It can serve arbitrarily many clients through a network of merge processes. Monitor queries use a programming technique called a \emph{hollow message}---a message that includes a writer, with which the consumer responds to the producer. Note that the program relaxes the SRSW requirement for consumed integers.
\begin{verbatim}
CounterCall::= add ; clear ; read(Integer?).

procedure monitor(Stream(CounterCall)?).
monitor(In) :- monitor(0,In?).

procedure monitor(Integer?,Stream(CounterCall)?).
monitor(N,[add|In]) :- N1 := N? +1, monitor(N1?,In?).
monitor(_,[clear|In]) :- monitor(0,In?).
monitor(N,[read(N?)|In]) :- integer(N?) | monitor(N?,In?).
\end{verbatim}

Additional examples demonstrating bounded-buffer communication via hollow messages, cooperative stream construction with interactive types, and abstract data types (difference lists and bidirectional channels) are in the repository, \verb|programs/TGLP/|.
\fi

\section{GLP Well-Typing}
\label{section:well-typing}
\subsection{Moded Paths}
We type a moded term (Definition~\ref{def:moded-term}) by relating the paths of its term tree to those of a GLP type.
\ifappendix
A moded term can be equivalently represented as a \emph{moded term-tree}, an LP term tree with edges labelled by (argument index, mode) pairs.  For example, the subtree for the first argument is:
\begin{verbatim}
(0,↓): merge/3
    (1,↓): "."/2
        (1,↓): 3
        (2,↓): Xs?
\end{verbatim}
with moded path \verb|(0,↓) --> merge/3 --(1,↓)--> "."/2 --(1,↓)--> 3|.
\else
A moded term has an equivalent \emph{moded term-tree} representation: a labelled tree with functors/arities at compound vertices, constants or variables at leaves, and edges labelled by (argument index, mode) pairs.
\fi

\begin{definition}[Moded Paths]
\label{def:moded-paths}
For a moded term $T$, $paths(T)$ denotes the set of \temph{moded paths} of its moded term tree.
A path is an \temph{input} (\temph{output}) path according to the mode on the edge leaving its root---the first-argument edge: consume~$\downarrow$ gives an input path, produce~$\uparrow$ an output path. The mode on the root functor itself is not used for this classification.  A moded term is \temph{consumed} (\temph{produced}) if all its mode annotations are consume (produce), and \temph{I/O} if it is $\downarrow$-annotated with at most one mode-inversion to $\uparrow$ on any of its paths.
\end{definition}

A defined GLP type $D$ also defines a regular set of moded paths, denoted $paths(D)$, of the same form, ending at a primitive type or wildcard. The type names $D$ passes through along a path are not themselves path positions. A moded term path, by contrast, ends at a primitive term---a constant or a variable.

\begin{definition}[Consistent Paths]
\label{def:consistent-paths}
Let $x$ be a moded term path and $y$ a GLP type path, with lengths $|x|$ and $|y|$ respectively.  Let $k = \min(|x|, |y|)$.  Then $x$ and $y$ are \temph{consistent} if:
\begin{enumerate}
    \item For all positions $i < k$ the functor/arity at position $i$ are identical in $x$ and $y$, and for all $i$ with $1 \le i < k$ the modes are identical; the mode at the root (position~0) is not compared.
    \item The symbols at position $k$ in $x$ and $y$ are \temph{compatible} per the table below.
\end{enumerate}
\end{definition}

\begin{center}
\begin{tabular}{|c|l|l|}
\hline
\textbf{\#} & \textbf{Term} & \textbf{Type} \\
\hline
1 & $f/n$, mode $m$ & $f/n$, mode $m$ \\
\hline
2 & \verb|X?| & \verb|_?|, or any symbol with mode $\downarrow$ \\
\hline
3 & \verb|X| & \verb|_|, or any symbol with mode $\uparrow$ \\
\hline
4 & integer & \verb|Integer|, \verb|Number|, or \verb|_| \\
\hline
5 & real & \verb|Real|, \verb|Number|, or \verb|_| \\
\hline
6 & string & \verb|String| or \verb|_| \\
\hline
7 & constant $c$ & $c$, \verb|String|, or \verb|_| \\
\hline
8 & module term & \verb|Module| \\
\hline
9 & any, mode $\downarrow$ & \verb|_?| \\
\hline
10 & any, mode $\uparrow$ & \verb|_| \\
\hline
\end{tabular}
\end{center}
The last two rows apply when the type path ends at a wildcard; the remainder of the term path is not examined.  The wildcard states \verb|_| and \verb|_?| subsume all primitive types of the matching mode.  Rows~2 and~3 handle the dual case: the term path ends at a variable while the type path continues, and the remaining type structure is not examined.

\ifappendix
\begin{example}[Consistent Paths]
Consider the first \verb|merge| clause with type \verb|merge(Stream?,Stream?,Stream)|.  For the first argument, the type path\newline \verb|(0,↓) --> merge --(1,↓)--> Stream? --(1,↓)--> "."/2 --(1,↓)--> _?| 
and term path\newline
\verb|(0,↓) --> merge --(1,↓)--> "."/2 --(1,↓)--> X?| are consistent: the paths are identical until position $k=4$, where the term has reader \verb|X?| and the type has \verb|_?|, which are compatible.
\end{example}
\fi

\begin{definition}[Well-Typed Moded Term]
A  moded term $T$ is \temph{well-typed} by a GLP type $D$ if for each term path $x\in paths(T)$ there is a consistent path $y \in paths(D)$ and the types according to $D$ of every variable pair in $T$ are dual.
\end{definition}

Clause heads require special treatment.  Three regimes of variable placement arise, all consistent with the writer-produces / reader-consumes convention.  In a goal or body term, a writer sits at each produced ($\uparrow$) position and a reader at each consumed ($\downarrow$) position.  A \emph{source clause head} inverts this: a head writer~$X$ at a $\downarrow$ position is an \emph{input slot}, and a head reader~\verb|X?| at an $\uparrow$ position is an \emph{output placeholder}.  A head is moreover I/O-moded: mode inversions occur at most once per path.  The moded head (Definition~\ref{def:moded-head}) restores the goal convention by complementing every head variable (step~2), so readers land at every $\downarrow$ and writers at every $\uparrow$---the form Definition~\ref{def:consistent-paths} checks.

\begin{definition}[Moded Head]
\label{def:moded-head}
Given a head $H$ for procedure $p$ with type declaration $p(T_1, \ldots, T_n)$, the \temph{moded head} $H'$ is constructed as follows:
\begin{enumerate}
    \item \textbf{Assign modes:} For each argument position $i$, assign mode $\downarrow$ if $T_i$ is an input type (written with trailing \verb|?|) and mode $\uparrow$ if $T_i$ is an output type.  Modes propagate through nested term structure according to the type definition, complementing at embedded \verb|?| annotations.
    \item \textbf{Complement variables:} Replace each variable with its paired variable.
\end{enumerate}
\end{definition}

\begin{example}[Moded Head]
Consider the first \verb|merge| clause head
\begin{verbatim}
H = merge([X|Xs], Ys, [X?|Zs?])
\end{verbatim}
with type \verb|merge(Stream?,Stream?,Stream)|.  Arguments 1 and 2 are consumed (\verb|Stream?|), argument 3 is produced (\verb|Stream|), yielding the I/O-moded term:
\begin{verbatim}
↓merge(↓[↓X|Xs], Ys, ↑[↑X?|Zs?])
\end{verbatim}
Complementing each variable yields the moded head:
\begin{verbatim}
H' = ↓merge(↓[↓X?|Xs?], Ys?, ↑[↑X|Zs])
\end{verbatim}
\end{example}

Without step~2 (variable complementation), the resulting term \verb!↓merge(!\allowbreak\verb!↓[↓X|Xs],! \verb!Ys,! \verb!↑[↑X?|Zs?])! places a writer~\verb|X| at a consumed position and a reader~\verb|X?| at a produced position.

\ifappendix
\begin{remark}[Mode Correspondence]
\label{rem:mode-correspondence}
Let $H'$ be a moded head constructed per Definition~\ref{def:moded-head} from head $H$ and type declaration $D$.  For any term path $x \in paths(H')$ ending at position $p$, and corresponding type path $y \in paths(D)$ at the same position $p$, the mode at $p$ in $x$ equals the mode at $p$ in $y$.

The same property holds for body unit goals: each body unit goal $A$ is a produced moded term---mode~$\uparrow$ at the root, with consumed embedded positions at~$\downarrow$---matching the type's expectation that body unit goals are produced; the root mode is not compared (Definition~\ref{def:consistent-paths}).

Consequently, when checking path consistency (Definition~\ref{def:consistent-paths}) for a moded head or body unit goal, the ``mode of the corresponding type symbol'' can be read directly from the term path.
\end{remark}
\fi

\subsection{Type Automaton}
\label{sec:type-automaton}

A GLP type definition $D$ (whether for a type $T$ or its dual $T?$) induces a \emph{type automaton}---a deterministic finite automaton that recognises the regular set of moded paths $paths(D)$.  States correspond to types (both $T$ and its dual $T?$), procedures ($p/n$), and primitives.  Transitions are labelled with functor, arity, argument position, and mode.

To simplify type checking, we require type definitions to be deterministic: alternatives must be distinguishable by their top-level functor.\ifappendix  The formal automaton construction appears in Appendix~\ref{app:type-automaton}.\fi

\subsection{Well-Typing}

With these notions, we can define when a clause is well-typed.

\begin{definition}[Type Assignment]
\label{def:type-assignment}
For a variable $V$ occurring at position $p$ in a clause $C$ with type declaration $D$, the \temph{type of $V$} is the state reached in the type automaton for $D$ by the path from the root to $p$.
\end{definition}

\begin{definition}[Well-Typed Clause]
\label{def:well-typed-clause}
Let $C= (H :- B)$ be a GLP clause and $D$ a GLP type for all its procedures. 
Then $C$ is \temph{well-typed} by $D$ if:
\begin{enumerate}
    \item There is a moded head $H'$ corresponding to $H$ that is well-typed by $D$.
    \item For each unit goal $A\in B$, the produced moded term $A'$ corresponding to $A$ is well-typed by $D$.
    \item For every variable pair $X$ and $X?$ in $C$:
    \begin{enumerate}
        \item[(a)] If both occur in the head, or both occur in the body, their types are dual.
        \item[(b)] If one occurs in the head and the other in the body, they have the same type.
    \end{enumerate}
\end{enumerate}
\end{definition}

\begin{definition}[Input-Accepting Clause]
\label{def:input-accepting-clause}
A clause $C$ with moded head $H'$ \temph{accepts} an input path $x \in paths(D)$ if $H'$ has a path consistent with $x$.
\end{definition}

For example, the input type path \verb|(0,↓) →| \verb|merge →| \verb|(1,↓) →| \verb|"."/2 →| \verb|(1,↓) →| \verb|_?| (a non-empty first stream) is accepted by the first \verb|merge| clause (whose moded head has term path \verb|(0,↓) →| \verb|merge →| \verb|(1,↓) →| \verb|"."/2 →| \verb|(1,↓) →| \verb|X?|, consistent by rows~1 and~2 of the table) and by the third clause (where \verb|X?| at position~1 matches any \verb|↓|-mode subtree).  It is not accepted by the fourth clause, whose first argument is the empty list \verb|[]| rather than \verb|"."/2|.

\begin{example}[Well-typed Clause]
We verify the first \verb|merge| clause is well-typed:
\begin{verbatim}
merge([X|Xs], Ys, [X?|Zs?]) :- merge(Ys?, Xs?, Zs).
\end{verbatim}

\emph{Condition 1 (Head well-typed):} The moded head from the example is:
\begin{verbatim}
H' = ↓merge(↓[↓X?|Xs?], Ys?, ↑[↑X|Zs])
\end{verbatim}
Each path in $H'$ is consistent with a path in $paths(D)$, as shown above.

\emph{Condition 2 (Body unit goals well-typed):} The body unit goal \verb|merge(Ys?, Xs?, Zs)| as a produced moded term \verb|↑merge(Ys?, Xs?, Zs)| has paths consistent with \verb|merge(Stream?,|\allowbreak\verb|Stream?,|\allowbreak\verb|Stream)|.

\emph{Condition 3 (Variable compatibility):} All four variable pairs satisfy the head-head/head-body dispositions of Definition~\ref{def:well-typed-clause} condition~3: \verb|X|/\verb|X?| are head-head duals; \verb|Xs|/\verb|Xs?|, \verb|Ys|/\verb|Ys?|, and \verb|Zs|/\verb|Zs?| are head-body same-type.
\end{example}

\ifappendix
\begin{example}[Moded Clauses for \texttt{merge}]
\label{ex:moded-merge-complete}
We illustrate moded clauses for the \verb|merge| program.  Each moded clause consists of a moded head $H'$ (constructed per Definition~\ref{def:moded-head}) and moded body unit goals (each a produced moded term).

\medskip
\noindent\textbf{Clause 1 (recursive):} \verb!merge([X|Xs], Ys, [X?|Zs?]) :- merge(Ys?, Xs?, Zs).!

\noindent Moded clause:
\begin{verbatim}
↓merge(↓[↓X?|Xs?], Ys?, ↑[↑X|Zs]) :- ↑merge(Ys?, Xs?, Zs).
\end{verbatim}
Variable types: \verb|X|:\verb|_|, \verb|X?|:\verb|_?| (head-head, duals); \verb|Xs|,\verb|Xs?|:\verb|Stream| (head-body, same); \verb|Ys|,\verb|Ys?|:\verb|Stream?| (head-body, same); \verb|Zs|,\verb|Zs?|:\verb|Stream| (head-body, same).

\medskip
\noindent\textbf{Clause 3 (base case):} \verb|merge(Xs, [], Xs?).|

\noindent Moded clause:
\begin{verbatim}
↓merge(Xs?, ↓[], ↑Xs).
\end{verbatim}
Variable types: \verb|Xs|:\verb|Stream|, \verb|Xs?|:\verb|Stream?| (head-head, duals).

\medskip
\noindent In each moded clause, the head is I/O-moded (arguments 1 and 2 consumed, argument 3 produced), and variables are replaced by their paired forms to match position modes.  All variable pairs satisfy type consistency: head-head pairs have dual types, head-body pairs have the same type.
\end{example}
\fi

\begin{definition}[Well-Typed GLP Program]
\label{def:well-typed-program}
A typed GLP program $P=(Cs,D)$ is \temph{well-typed} if:
\begin{enumerate}
    \item Every clause $C \in Cs$ is well-typed by $D$.
    \item Every input path in $D$ is accepted by some clause $C\in Cs$.
\end{enumerate}  
\end{definition}

\subsection{Subtyping}
\label{sec:subtyping}

Definition~\ref{def:well-typed-clause} requires that body-body variable pairs have dual types.  This requirement can be relaxed using \emph{subtyping}: it suffices that the type of the writer~$X$ be a subtype of the dual of the type of its paired reader~$X?$.  Informally, anything the writer \verb|X| produces, the reader \verb|X?| can consume.  Where a subterm reverses the direction of communication, the subtyping direction reverses with it.

A file system monitor handling \verb|read|, \verb|write|, and \verb|delete| operations has type:
\begin{verbatim}
FileOp ::= read(Path?,Content)
         ; write(Path?,Content?)
         ; delete(Path?).
\end{verbatim}
\ifappendix with procedure declaration \verb|procedure fs_monitor(Stream(FileOp)?)| (where \verb|Stream(T)| denotes a parameterised stream of \verb|T| values; see Section~\ref{sec:parameterized-types}).\fi
A read-only client that only issues \verb|read| operations has the narrower type \verb|ReadOp ::= read(Path?,Content)|\ifappendix, with declaration \verb|procedure ro_client(Stream(ReadOp))|\fi.  Connecting them directly violates strict duality (\verb|ReadOp| $\neq$ \verb|FileOp|), yet the interaction is safe: the client produces a subset of what the monitor accepts, and the monitor's responses (the \verb|Content| in \verb|read|) are exactly what the client expects.

\begin{definition}[Simple Prefix]
\label{def:simple-prefix}
A \temph{simple prefix} of an output type $T$ is a path in $T$'s type automaton starting from $T$ that contains no mode inversions.  A simple prefix ends when it reaches either:
\begin{itemize}
    \item The produced primitive \verb|_|, or
    \item A mode inversion point: any position whose type is marked with \verb|?|.
\end{itemize}
\end{definition}

\begin{definition}[Prefix Acceptance]
\label{def:prefix-acceptance}
A simple prefix $p$ of type $A$ is \temph{accepted by} type $B$ if $B$ has a simple prefix $q$ with identical functor/position structure, where endpoints satisfy: \verb|_| matches only \verb|_|; output type $S$ matches $S$ or \verb|_|; a constant matches \verb|String|; mode inversion $S?$ matches any mode inversion at the same position.
\end{definition}

\begin{definition}[Subtyping]
\label{def:subtyping}
Let $A$ and $B$ be GLP output types.  We say $A$ is a \temph{subtype of} $B$, written $A <: B$, if:
\begin{enumerate}
    \item Every simple prefix of $A$ is accepted by $B$.
    \item For every mode inversion point in $A$ reached by a simple prefix---say, type $A'?$ at that position---there is a corresponding mode inversion point $B'?$ in $B$ at the same position, and $B'$ is a subtype of $A'$.
\end{enumerate}
$<:$ is the greatest relation satisfying conditions 1 and 2.
\end{definition}

Decidability follows from the finiteness of the type DFA: $<:$ is checked by a finite simulation between the DFAs of $A$ and $B$.

\ifappendix
\begin{example}[Subtyping Analysis]
We verify that \verb|ReadOp| is a subtype of \verb|FileOp|.  \verb|ReadOp| has functor \verb|read/2| only, while \verb|FileOp| includes \verb|read|, \verb|write|, and \verb|delete|.

\emph{Condition 1:} Each simple prefix of \verb|ReadOp| is accepted by \verb|FileOp|: position 1 reaches \verb|Path?| (mode inversion) in both types; position 2 reaches \verb|Content| (output) in both.

\emph{Condition 2:} At the mode inversion point (\verb|Path?|), both types have identical \verb|Path|, so the contravariant check holds trivially.

Thus \verb|ReadOp| $<:$ \verb|FileOp|, and a writer of type \verb|ReadOp| paired with a reader of type \verb|FileOp?| is well-typed.
\end{example}
\fi

\ifappendix Subtyping generalises exact duality: if $S = T$ then $S <: T$.  The relation is transitive but not symmetric: \verb|ReadOp| $<:$ \verb|FileOp| but not conversely.\fi

\begin{definition}[Well-Typed Clause with Subtyping]
\label{def:well-typed-clause-subtyping}
A GLP clause is \temph{well-typed with subtyping} if it satisfies Definition~\ref{def:well-typed-clause} with condition~3(a) relaxed for body-body variable pairs, requiring that the type of the writer be a subtype of the dual of the type of its paired reader.
\end{definition}

\begin{definition}[Well-Typed GLP Program with Subtyping]
\label{def:well-typed-program-subtyping}
A typed GLP program $P=(Cs,D)$ is \temph{well-typed with subtyping} if it satisfies Definition~\ref{def:well-typed-program} with ``well-typed clause'' replaced by ``well-typed clause with subtyping'' (Definition~\ref{def:well-typed-clause-subtyping}).
\end{definition}

\begin{remark}[Subtyping Variance]
\label{rem:subtyping-variance}
The coinductive structure of Definition~\ref{def:subtyping} embodies the standard variance principle: covariant in output positions (the subtype's functors must be contained in the supertype's), contravariant in input positions (at mode inversion points, the containment direction reverses).  This matches the variance of session type subtyping~\cite{gay2005subtyping}.
\end{remark}

\section{GLP Semantics and Well-Typing}
\label{sec:glp-denotational}
\label{sec:glp-semantics-well-typing}

\ifappendix Similarly to LP, we\else We\fi\ derive denotational semantic notions for which the type system is an abstraction.  The semantic objects we employ---moded terms---are richer than in LP: they capture both partial-run behaviour (so well-typing constrains runs that may fail, deadlock, or never terminate) and the produced/consumed distinction (supporting both output conformance and input coverage).

For a procedure declaration \verb|p(|$T_1?, \ldots, T_k?, T_{k+1}, \ldots, T_n$\verb|)|, input paths traverse the consumed arguments $T_1?, \ldots, T_k?$ and output paths traverse the produced arguments $T_{k+1}, \ldots, T_n$.

\subsection{Well-Typed Outcomes}

The moded-atoms outcome and moded resolvent of a run (Definition~\ref{def:moded-outcome}) are \temph{well-typed} by $D$ if the moded terms $G_0'$ and $G'$, respectively, are well-typed by $D$.  A goal $G_0$ is \temph{well-typed} by $D$ if it is well-typed as a body.  Note that $G_0$ can be conjunctive, but there is only a single program $D$.  The \temph{well-typed moded-atom semantics} of a typed GLP program $P=(Cs,D)$ is the set of all moded-atoms outcomes of every finite proper run of $P$ with a well-typed initial goal.

\subsection{Soundness of Well-Typing}

Next we connect syntactic well-typing with the semantic \emph{output conformance} of runs from well-typed initial goals, and prove it sound.

\begin{definition}[Output Conformance]
\label{def:output-conformance}
An output moded path $x$ \temph{conforms to} a GLP type $D$ if either
\begin{enumerate}
\item[(\ia)] $x$ is consistent with some path in $paths(D)$ (Definition~\ref{def:consistent-paths}); or
\item[(\ib)] $x$ ends at a reader, and deleting that reader leaves a prefix agreeing in functor, arity, and non-root mode with a proper prefix of some path in $paths(D)$.
\end{enumerate}
\end{definition}

A reader at a produced position is the consumer's handle on an output the run has not yet produced; its producing writer occurs in the resolvent.  Case~(\ib) admits such a pending output.  It does not relax Definition~\ref{def:consistent-paths}, which stays strict and still rejects a reader at a produced position in a clause head---a mode error---because clause well-typing is defined by consistency, not conformance.

\begin{lemma}[Well-Typing Preservation]
\label{lem:well-typing-preservation}
Let $P=(Cs,D)$ be well-typed and $\rho:(G_0,\emptyset)\to\cdots\to(G_n,\sigma_n)$ a proper run from a well-typed initial goal $G_0$.  Then for every $i\le n$:
\begin{enumerate}
\item every input path of the moded-atoms outcome $G_0\sigma_i^\star$ is consistent with $D$, and every output path of it conforms to $D$ (Definition~\ref{def:output-conformance});
\item the moded resolvent $G_i\sigma_i^\star$ is well-typed by $D$ as a body; and
\item every reader at a produced position of $G_0\sigma_i^\star$ is paired with a writer of $G_i\sigma_i^\star$ of the same type under $D$.
\end{enumerate}
\end{lemma}

\ifappendix
\begin{proof}
By induction on $i$.

\emph{Base} ($i=0$): $\sigma_0=\emptyset$, so outcome and resolvent are both $G_0'$.  As a well-typed initial goal, $G_0'$ is well-typed by $D$ as a body, giving (1) and (2); its output positions carry writers, not readers, so (3) holds vacuously.

\emph{Communicate} leaves $\sigma$ unchanged, hence leaves the outcome unchanged, so (1) and (3) carry over.  For (2), a reader $X?\in G_{i-1}$ is replaced by $T$ with $\{X?:=T\}\in\sigma_{i-1}$; the assignment $X:=T$ was made by an earlier Reduce, so $T$ conforms to the type of $X$, and by (3) at that step $X$ and $X?$ have the same type, so $T$ conforms at $X?$'s position.

\emph{Reduce.}  Unit goal $A\in G_{i-1}$ reduces with the first applicable clause $C=(H\mathrel{\texttt{:-}}B)$, renamed apart, via writer mgu $\hat\sigma$; $G_i=(G_{i-1}\setminus\{A\}\cup B)\hat\sigma$ and $\sigma_i=\sigma_{i-1}\circ\hat\sigma?$.  Since $P$ is well-typed, $C$ satisfies the three conditions of Definition~\ref{def:well-typed-clause}.

Matching $A$ against $H$ (Definition~\ref{def:term-matching}) succeeds, so every joint position agrees; by the writer-produces/reader-consumes convention and Remark~\ref{rem:mode-correspondence} the matched positions carry equal modes in $A'$ and $H'$.  Each assignment of $\hat\sigma$ is thus either (a)~at a produced position, a writer of $A$ bound to the output structure of $H$ there, identifying $A$'s pending output with what the clause produces; or (b)~at a consumed position, an input slot of $H$ bound to the subterm or reader $A$ supplies.  In case~(a) the head structure's leaves are output placeholders, each the head reader of a head--body pair whose body writer has the same type by condition~(3); in case~(b) the bound subterm is $A$'s input, consistent with $D$ by the induction hypothesis applied to $A$.

\emph{(2).}  By condition~(2) each goal of $B$ is a produced moded term well-typed by $D$; $\hat\sigma$ fills its input positions with $A$'s (consistent) input and leaves its output positions as writers, so each goal of $B\hat\sigma$ has consistent input and conforming output.  The goals of $G_{i-1}\setminus\{A\}$ are renamed apart from $C$ and unchanged.  Hence $G_i\sigma_i^\star$ is well-typed as a body.

\emph{(1) and (3).}  The outcome changes from $G_0\sigma_{i-1}^\star$ only through $\hat\sigma?$, the reader counterparts of the writers $\hat\sigma$ binds.  The clause's own variables are fresh, so the only counterparts reaching the outcome are those of $A$'s output writers (case~(a)); by the induction hypothesis~(3) each such writer is paired with a pending reader of the outcome.  Binding it refines that reader to the clause's produced structure at the position---which conforms by condition~(1)/(2)---with the body's writers leaving fresh pending readers below it.  By condition~(3) each fresh pending reader has the same type as its body writer, now a goal of $G_i$, re-establishing (3).  Every refined output position therefore conforms (a consistent prefix ending in clause-produced structure or, by case~(\ib), in a fresh pending reader), and input positions are unchanged or refined by consistent input, giving (1).
\end{proof}
\fi

\begin{theorem}[Soundness of Well-Typing]
\label{thm:soundness}
If a typed GLP program $P = (Cs, D)$ is well-typed, then every output path in its well-typed moded-atom semantics conforms to $D$ (Definition~\ref{def:output-conformance}).
\end{theorem}

This is the covariant, or subsumption, direction. Well-typing soundly over-approximates the semantics: it guarantees type-conformant output, not freedom from failure or deadlock.

\ifappendix
\begin{proof}
By Lemma~\ref{lem:well-typing-preservation}(1), every output path of the moded-atoms outcome $G_0\sigma_i^\star$ of a run from a well-typed initial goal conforms to $D$; these outcomes constitute the well-typed moded-atom semantics.
\end{proof}
\fi

Soundness has a contravariant counterpart for inputs.

\begin{proposition}[Input Coverage]
\label{prop:input-coverage}
Let $P=(Cs,D)$ be well-typed. Then every input path in $paths(D)$ is accepted by some clause in $Cs$, and every input path of every clause's moded head is consistent with some path in $paths(D)$.
\end{proposition}

Soundness contains a program's outputs within $D$; coverage requires every input $D$ declares to be matched by some clause head. This mirrors session subtyping, covariant in outputs and contravariant in inputs~\cite{gay2005subtyping}. Soundness is a property of the moded-atom semantics, coverage of the clause heads. Coverage holds per input path, not per argument tuple, so a goal matching no clause may still fail.

\ifappendix
\begin{proof}
The first conjunct is condition~2 of Definition~\ref{def:well-typed-program}: a clause accepts an input path $x\in paths(D)$ when its moded head has a path consistent with $x$ (Definition~\ref{def:input-accepting-clause}). The second is condition~1 of Definition~\ref{def:well-typed-clause}: each moded head is well-typed by $D$, so each of its paths, in particular each input path, is consistent with some path in $paths(D)$. Equality of non-root modes (Definition~\ref{def:consistent-paths}) makes each consistent path an input path.
\end{proof}
\fi

\begin{corollary}[Extension to Subtyping]
\label{cor:subtyping-extension}
Theorem~\ref{thm:soundness} holds with ``well-typed'' replaced by ``well-typed with subtyping'' (Definition~\ref{def:well-typed-program-subtyping}).
\end{corollary}

\ifappendix
\begin{proof}
The relaxation touches only body-body variable pairs, where a writer $X$ of type $S'$ is paired with a reader $X?$ of type $S?$ and we require $S' <: S$ in place of $S' = S$.  Lemma~\ref{lem:well-typing-preservation} is unchanged except at such a pair, where its appeal to equal types becomes an appeal to subtyping.  By Definition~\ref{def:subtyping}---covariant on produced structure and contravariant at mode inversions (Remark~\ref{rem:subtyping-variance})---every value the writer produces conforms to $S$, which the reader's type $S?$ admits, the inclusions reversing at each embedded mode inversion as the dual roles require.  Clause~(3) of the lemma thus still pairs every reader at a produced position with a writer of a conforming type, and output conformance is preserved.  The theorem follows as before.
\end{proof}
\fi

%These results carry over to parameterised types, introduced next.

\ifappendix
%==============================================================================
\section{Module System}
\label{sec:modules}
%==============================================================================

This section extends Typed GLP with a module system for type-safe composition of separately developed modules.

\begin{definition}[Typed Procedure, Module]
\label{def:module}
A \temph{typed procedure} is a procedure declaration \verb|procedure |$p(T_1,\ldots,T_n)$\verb|.| immediately followed by a procedure for $p/n$ (\cref{def:glp-program}).  A \temph{module} is a sequence of type definitions and typed procedures, in which every type is defined before its use in a declaration, with a subset of the typed procedures \temph{exported}.
\end{definition}

\subsection{Design}
\label{sec:mod-design}

The module system rests on four principles.

\mypara{Hierarchy mirrors the file system}
A GLP program is organised as a source-code directory hierarchy, with a top-level \verb|self.glp| module as its \emph{root}.  Each \verb|.glp| file is a module; each directory is a scope.  A distinguished file \verb|self.glp| in any directory defines types and procedures shared by all modules in that directory and its descendants, following FCP's \verb|self.cp| convention~\cite{silverman1988logix}.  A directory's \verb|self.glp| is the directory's interface: alongside the shared definitions it declares which of the directory's procedures are \temph{exported}, and a call into the directory from outside it resolves only to a procedure its \verb|self.glp| exports.  A Typed GLP module is \temph{self-contained} if it does not include a cross-module call \verb|M#p| or an uninstantiated type parameter.  A Typed GLP program is either a self-contained module or a directory with a \verb|self.glp| module.  Programs are linked within their filesystem context, type-checked, compiled (if they pass type checking), and run.

\mypara{Implicit ancestor scoping}
A module sees every type and procedure definition from every ancestor scope without any import directive.  The chain is anchored at the hierarchy root, not at the directory loaded for execution.

\mypara{Self-contained type checking}
Every module declares the full moded type of every cross-module procedure it calls, via \verb|imported procedure| declarations.  A module may be checked separately against its own text and its ancestor scope; soundness is established on the linked program (\cref{def:program}), the unit of type checking.

\mypara{Structural type compatibility}
Two types with the same type automaton are compatible, regardless of their names or defining modules.

\mypara{Procedure declarations}
Every procedure declaration is one of three kinds:

\begin{center}
\begin{tabular}{ll}
\verb|procedure p(...).| & private to the module \\
\verb|exported procedure p(...).| & callable from other modules \\
\verb|imported procedure M#p(...).| & declares dependency on module~$M$ \\
\end{tabular}
\end{center}

Types referenced in an \verb|imported| declaration are resolved against the caller's type scope.  Export and import are declared per procedure; there are no separate export or import lists.  A declaration carries the transitive closure of the types its signature references, so types are not exported separately.

\mypara{External access}
A procedure of a program is externally accessible only if the root \verb|self.glp| exports it.  The root \verb|self.glp| exports a procedure by declaring it \verb|exported| and either defining it directly or giving a forwarding clause whose body is the cross-module call $M\#p$ to a module that defines $p$, each argument threaded at its declared polarity.  Depth does not decide accessibility: a procedure is an entry point exactly when the root \verb|self.glp| exports it, by definition or by forwarding.

\mypara{Scope construction}
The scope for a module $M$ in a hierarchy rooted at $r$ begins with the GLP language primitives (Appendix~\ref{sec:root-self}).  For each directory $d_1, d_2, \ldots, d_k$ on the path from $r$ to the directory containing $M$, with $d_1 = r$, if $d_i$ contains a \verb|self.glp| its type definitions and procedure declarations are added, later definitions shadowing earlier ones; finally the type and procedure definitions of $M$ itself are added.

Common protocol types are defined once at the lowest common ancestor of all modules that use them.

\mypara{The \texttt{-expose} directive}
A \verb|self.glp| may contain the directive \verb|-expose(M).|, which lifts the exported procedures of module \verb|M| (and the types their signatures carry) into that directory's scope, as if defined in its \verb|self.glp|.  \verb|M| is a module path (e.g., \verb|lib#streams|), resolved relative to the directory of the \verb|self.glp| containing the directive, within that directory's subtree.  Shadowing applies as usual; two exposed modules contributing the same name and arity at one level is an error.  An exposed procedure is callable by name inside the directory and its descendants, but it is not thereby exported by the root \verb|self.glp|, so it is an entry point only if the root \verb|self.glp| exports it in its own right.

\mypara{Cross-module type checking}
A cross-module call \verb|M # p(|$a_1, \ldots, a_n$\verb|)| in module~$N$ is well-typed if $N$ contains a declaration \verb|imported procedure M#p(|$T_1, \ldots, T_n$\verb|)| and for each argument $a_i$, the type of $a_i$ in the clause context is compatible with $T_i$ under the standard well-typing rules of Section~\ref{section:well-typing}, including subtyping with mode variance.  This is identical to the well-typing rule for local calls, substituting the imported declaration for a local one.  The qualifier \verb|M| is a single child directory or module file relative to the caller's directory: a directory is entered through its \verb|self.glp|, a module file through its own \verb|exported| declarations.  Multi-segment path qualifiers \verb|d1#...#dk#p| are future work.

\mypara{Example}
Consider a social graph application with an agent module and a boot module.  Shared types are defined in \verb|self.glp|:

\begin{verbatim}
%% self.glp
Response ::= accept(FriendChannel) ; no.
AgentContent ::= befriend(Constant, Response?)
               ; connected(Constant)
               ; rejected.
AgentChannel ::= ch(AgentToUserStream, UserToAgentStream?).
\end{verbatim}

The agent module exports a single procedure:

\begin{verbatim}
%% agent.glp
exported procedure agent(Constant?, UserInStream?,
                         NetInStream?, OutputsList?).
procedure merge(Stream(X)?, Stream(X)?, Stream(X)).
\end{verbatim}

The boot module declares its dependency:

\begin{verbatim}
%% boot.glp
imported procedure agent#agent(Constant?, UserInStream?,
                               NetInStream?, OutputsList?).
\end{verbatim}

and calls \verb|agent # agent(alice, UserIn?, NetIn?, Outs)|.  The type checker verifies this call against the local \verb|imported| declaration using the types from \verb|self.glp| (visible to both modules via ancestor scoping), without reading \verb|agent.glp|.  Both modules may define a private \verb|merge/3| without conflict, since each is scoped to its own module.

\subsection{Static Linking}
\label{sec:static-linking}

When all modules are available at compile time, \emph{static linking} transforms the module hierarchy into a single flat program, compiled through the standard pipeline.  All inter-module calls become local calls.

The transformation proceeds in five steps.  First, the compiler walks the program's directory tree, collecting every \verb|.glp| file and constructing the ancestor scope chain for each module.  Second, each module is type-checked independently against its ancestor scope, exactly as for single-file compilation.  Third, after type checking, every procedure $p/n$ in every \verb|.glp| file (including \verb|self.glp| files) is renamed to $M{:}p/n$, where $M$ is the module's path from the program root, eliminating name collisions---for example, both \verb|agent.glp| and \verb|boot.glp| may define \verb|merge/3|, yielding \verb|agent:merge/3| and \verb|boot:merge/3|, while \verb|secure/boot.glp| yields the distinct \verb|secure/boot:merge/3|.  Fourth, every goal in every clause body is resolved: a local call to $p$ in module~$M$ becomes $M{:}p$, a cross-module call $M'\#p$, whose qualifier is a single child directory or module file relative to the caller's directory, resolves to the procedure $p$ that the qualifier exports---a directory exports through its \verb|self.glp|, a module file through its own \verb|exported| declarations---renamed to its prefix ($M'{:}p$), and a call matching a procedure in an ancestor \verb|self.glp| is resolved to its renamed form by walking the ancestor chain.  Exposed procedures resolve like procedures of the exposing directory's \verb|self.glp|.  Fifth, the linker passes to the compiler the \temph{reachable} procedures: the exported procedures of the root \verb|self.glp|, and every procedure called in the body of a reachable one.  These exports are the program's entry points, called by plain name by a goal posted at the root.  A single-module program, having no \verb|self.glp|, exports all its procedures, so every one is an entry point.  Restricting the program to its reachable procedures is semantically equivalent to the whole.

The flat program is the linked program of \cref{def:program}, and type checking it verifies each procedure's clauses at the concrete types of every call to it, now that linking has made every call local.  This check is more stringent than the per-module check, as it checks the called procedure's clauses against the concrete types the call supplies.

Only a well-typed program (\cref{def:well-typed-program-modular}) is compiled and run.

\subsection{Dynamic Activation}
\label{sec:dynamic-linking}

Dynamic activation runs a goal on a module that is not present at link time but arrives at runtime as a value of the primitive type \verb|Module|.  Its type discipline rests on a \emph{type identity}: the hash of the full type automaton of a procedure declaration, in canonical minimised form, with every type the signature references expanded---not of the declaration's text.  By structural type compatibility (Section~\ref{sec:mod-design}), two declarations are compatible exactly when their automata are equal, hence exactly when their type identities are equal.  Two primitives provide the identity and the activation:

\begin{itemize}
\item \verb|find_type(P/N, Type)| assigns to \verb|Type| the type identity of the declaration of procedure \verb|P/N| in the caller's scope, and \verb|error| if \verb|P/N| is not declared there.
\item \verb|run(Goal, Type, Module)| activates \verb|Module| and posts \verb|Goal| to it if \verb|Type| equals the type identity that \verb|Module|'s exports record for \verb|Goal|'s predicate, and errs otherwise.
\end{itemize}

\noindent The activation check compares hashes; no runtime type checking of terms is introduced.  Soundness is conditional: if \verb|Goal| is well-typed by a declaration whose type identity is \verb|Type|, and \verb|run| activates, then the goal's author and the module hold equal automata for the conversation, and the terms crossing the boundary conform on both sides.  The type checker enforces the condition where the goal is authored: a clause that constructs \verb|Goal| and runs it must have \verb|Goal| well-typed by the declaration of its predicate in scope, with \verb|Type| that declaration's identity, obtained via \verb|find_type|.  Under delegation the running module holds no declaration of \verb|Goal|'s predicate: \verb|Goal| and \verb|Type| arrive together over a typed stream, the condition having been enforced where they were authored.

A procedure declared at a scope shared by the goal's author and the module---their lowest common ancestor \verb|self.glp|---has the same automaton on both sides by construction; the hash covers the case with no shared scope, a module shipped from another agent.  Within one agent the loader has validated every module, and equality of type identities completes the check; for a shipped module, that it is a validated, well-typed module whose exports record true identities rests on the attestation of Section~\ref{sec:runtime-boundary}.  The type identity refines the source hash used there: equal sources define equal automata, and the automaton hash admits equal automata defined by different sources, per structural compatibility.

Implementation of dynamic activation is future work; static linking (Section~\ref{sec:static-linking}) is the model in force.

\subsection{Type-Compatible Attestation Between Agents}
\label{sec:runtime-boundary}

The guarantee of Theorem~\ref{thm:soundness} requires a well-typed program and a well-typed initial goal.  Within a single program both are established at compile time.  A term arriving from another agent, however, is produced outside the scope of the receiver's type checker, so for such a term neither hypothesis is discharged by the checks above.

Cross-agent communication takes place only between mutually attested runtimes.  Participants prove to one another that they run the same module encoding the shared protocol, and that the variables through which they communicate carry the same types, by exchanging the module identity---the hash of its source---and the type identity of the conversation's root channel---the pair of the hash of the type's defining source and the type's name---and proceeding only on equality; derived variables undergo no handshake, their types being determined at both ends by the root type and the communications.  Equal sources define equal type automata, so the peer writes every variable of the conversation as a well-typed program under the same type discipline, discharging the well-typed-input hypothesis and eschewing runtime typechecking.  The architecture, and the theorem that mutually attested well-typed agents compose into a well-typed system, are developed in the companion paper~\cite{keidar2026secure}.

The remaining producer of unchecked terms is the initial goal posted to the runtime, at boot or interactively; it is type-checked before execution as a body goal (\cref{def:well-typed-clause}).

\subsection{Implementation}
\label{sec:mod-implementation}

The module system has been implemented in Dart as part of the GLP runtime.  Ancestor type scoping, the three procedure declaration kinds, and cross-module type checking are operational.

Static linking is implemented as the AST-to-AST transformation of Section~\ref{sec:static-linking}.

An earlier dynamic-dispatch mechanism, which routed cross-module calls through channels at runtime, remains present in the runtime but is retired: it cannot be typed without runtime type-checking, which the language avoids.  It will be removed once dynamic activation (Section~\ref{sec:dynamic-linking}) is implemented.

The system was validated on three applications restructured from monolithic programs into modular ones: a Child-Safe Social Graph with four agents and seven test scenarios, a simulated-UI variant, and a Child-Safe Social Network (CSSN) with twelve test scenarios covering group formation and messaging.  Each application runs correctly through static linking.

Before parameterised types, the \verb|boot| module in the Child-Safe Social Graph program could not be fully typed: its generic stream utilities (\verb|tee|, \verb|merge|, \verb|sink|) could not express ``this operation preserves the specific type of its input.''  With the parameterised type declarations of Section~\ref{sec:parameterized-types}, these utilities are now declared as \verb|procedure tee(Stream(X)?, Stream(X), Stream(X)).| and \verb|procedure merge(Stream(X)?, Stream(X)?, Stream(X)).|, and all boot modules are fully type-checked.

\fi
%==============================================================================
\section{Parameterised Types}
\label{sec:parameterized-types}
%==============================================================================

Monomorphic \verb|Stream| (carrying elements of type~\verb|_|) cannot express that \verb|merge| preserves the element type from inputs to output; parameterised types recover this precision.
Here, parameterised types are syntactic sugar: each use with concrete type arguments expands to a fresh monomorphic type definition before type automaton construction, after which existing well-typing and subtyping machinery apply unchanged.

\subsection{Parameterised Type Definitions}
\label{sec:param-syntax}
A parameterised type definition introduces type parameters, written as uppercase identifiers in parentheses after the type name:
\begin{verbatim}
Stream(X) ::= [] ; [X | Stream(X)].
\end{verbatim}
This defines a type \emph{template}.  An \emph{instantiation} supplies a concrete type for each parameter:

\begin{verbatim}
Stream(Integer)     % stream of integers
Stream(AgentMsg)    % stream of agent messages
\end{verbatim}

\ifappendix
Multiple parameters are supported:

\begin{verbatim}
Pair(A, B) ::= pair(A, B).
Channel(In, Out) ::= ch(In, Out?).
\end{verbatim}

Mode annotations within a template (\verb|Out?|) are preserved during expansion.
\fi

\begin{definition}[Parameterised Type Definition]
\label{def:parameterized-type}
A \temph{parameterised type definition} has the form $T(X_1, \ldots, X_k) ::= A_1 ; \ldots ; A_n$, where the \temph{type parameters} $X_1, \ldots, X_k$ may occur in place of type names within $A_1, \ldots, A_n$.  
\end{definition}

Where a parameterised type refers to itself, directly or transitively, no parameter may occur as a proper subterm of an argument of the self-referential occurrence.  Each type instantiation then expands to finitely many type definitions.

\subsection{Expansion}
\label{sec:param-expansion}

\mypara{Expansion rule}
Given a parameterised type definition $T(X_1, \ldots, X_k) ::= A_1 ; \ldots ; A_n$ and an instantiation $T(S_1, \ldots, S_k)$, the expansion is:
\[
T\langle S_1, \ldots, S_k \rangle ::= A_1[S_1/X_1, \ldots, S_k/X_k] \;;\; \ldots \;;\; A_n[S_1/X_1, \ldots, S_k/X_k]
\]
where $T\langle S_1, \ldots, S_k \rangle$ is a fresh type name and $A_i[S_j/X_j]$ denotes simultaneous substitution.  Every occurrence of $T(X_1, \ldots, X_k)$ in the body is replaced by $T\langle S_1, \ldots, S_k \rangle$.

\mypara{Example: \texttt{Stream(Integer)}}
From the template \verb!Stream(X) ::=! \verb![] ;! \verb![X |! \verb!Stream(X)].!, the instantiation \verb|Stream(Integer)| expands to:
\begin{verbatim}
Stream<Integer> ::= [] ; [Integer | Stream<Integer>].
\end{verbatim}
Its type automaton is constructed by the rules of Section~\ref{sec:type-automaton}, and well-typing proceeds as in Section~\ref{section:well-typing}.
\ifappendix
\mypara{Example: \texttt{Channel(FriendMsg, FriendMsg)}}
From \verb|Channel(In, Out) ::= ch(In, Out?)|, the instantiation \verb|Channel(FriendMsg, FriendMsg)| expands to:

\begin{verbatim}
Channel<FriendMsg,FriendMsg> ::= ch(FriendMsg, FriendMsg?).
\end{verbatim}

The channel's second component is an input stream from the holder's perspective.

\mypara{Nested parameterised types}
Parameters may themselves be parameterised type instantiations:

\begin{verbatim}
Stream(Pair(Integer, String))
\end{verbatim}

Expansion proceeds inside-out: first expand \verb|Pair(Integer, String)| to \verb|Pair<Integer,String> ::= pair(Integer, String).|, then expand \verb|Stream(Pair<Integer,String>)| to:

\begin{verbatim}
Stream<Pair<Integer,String>> ::= [] ; [Pair<Integer,String>
                                      | Stream<Pair<Integer,String>>].
\end{verbatim}
\fi
As expansion preserves the regular-set-of-moded-paths structure of types, the soundness of well-typing (\cref{thm:soundness}) extends to parameterised types.

\subsection{Parameterised Procedure Declarations}
\label{sec:param-procedures}

A procedure declaration may use type parameters to express uniform behaviour:

\begin{verbatim}
procedure merge(Stream(X)?, Stream(X)?, Stream(X)).
\end{verbatim}

The parameter \verb|X| is implicitly universally quantified.  At a call, each parameter or parameterised type in the declaration is matched against the concrete type at the call site, binding each parameter; conflicting bindings for the same parameter are a type error.  The declaration is a \emph{clause template}.  A call in a clause body to a parameterised procedure induces an instantiation of the callee: the caller's bindings make the call's argument types concrete, and matching them against the callee's declaration binds its parameters.  The \emph{instantiations} of a program form the least set containing the instantiation of every call in a monomorphic clause and closed under this induction, except that recursion is monomorphic: a call to a procedure already being instantiated on the current cycle is checked at that instantiation rather than inducing a new one, and a call that would require a different instantiation of such a procedure is a type error.  Recursion thus induces no instantiation, and the non-recursive calls induce finitely many, so the set is finite.  For each instantiation in the set, the clauses of the instantiated procedure must be well-typed (Definition~\ref{def:well-typed-clause}) by the monomorphic declaration the instantiation produces; \cref{sec:programs-and-modules} casts this as the well-typing of programs.  Every procedure instantiation a run reaches lies in the set, so soundness (\cref{thm:soundness}) extends to parameterised procedures.

\mypara{Declaration parameters}
The parameters of a procedure declaration are exactly the undefined names occurring within the arguments of template instantiations in it: \verb|X| is a parameter of \verb|merge| through its occurrence in \verb|Stream(X)|, and may also occur bare elsewhere in the declaration.  Any other undefined name in a declaration is an error, so a misspelt type name is rejected rather than read as a parameter.

\ifappendix
\mypara{Example: instantiation from call context}
Consider a typed program with:

\begin{verbatim}
AgentMsg ::= befriend(Constant, Response?)
           ; connected(Constant)
           ; rejected.

procedure agent_merge(Stream(AgentMsg)?,
                      Stream(AgentMsg)?,
                      Stream(AgentMsg)).
agent_merge(A, B, C) :- merge(A?, B?, C).
\end{verbatim}

The call \verb|merge(A?, B?, C)| occurs where \verb|A|, \verb|B|, \verb|C| have type \verb|Stream(AgentMsg)|.  The type checker matches \verb|Stream(X)| against \verb|Stream(AgentMsg)| and infers \verb|X = AgentMsg|.  It then expands the declaration to:

\begin{verbatim}
procedure merge(Stream<AgentMsg>?, Stream<AgentMsg>?, Stream<AgentMsg>).
\end{verbatim}

and type-checks the clause body against this monomorphic declaration.

\mypara{Channel operations}
The root \verb|self.glp| declares channel operations using the two-parameter \verb|Channel(In, Out)| template:

\begin{verbatim}
procedure send(X?, Channel(Y, Stream(X))?,
               Channel(Y, Stream(X))).
procedure receive(X, Channel(OpenStream(X), Y)?,
                  Channel(Stream(X), Y)).
procedure new_channel(Channel(X, Y), Channel(Y, X)).
\end{verbatim}

In \verb|send|, \verb|X| is the message element type and \verb|Y| the read stream type; the type checker infers both from the call context.  \verb|receive| consumes a message from a non-empty read stream (\verb|OpenStream(X)|); the defined guard \verb|close| covers the closed one (\cref{sec:root-self}).

The clause \verb|new_channel(ch(Xs?, Ys), ch(Ys?, Xs)).| produces two channel endpoints: what one end writes, the other reads, and vice versa.

\mypara{Multiple parameters}
\begin{verbatim}
procedure relay(Stream(X)?, Stream(X), Channel(X, X)?).
\end{verbatim}

If the context provides \verb|Stream(Integer)?| for the first argument and \verb|Channel(String, String)?| for the third, the type checker reports an error: \verb|X| cannot be both \verb|Integer| and \verb|String|.
\fi

\subsection{Programs and Modules}
\label{sec:programs-and-modules}

Parameterised type definitions and parameterised procedure declarations are syntactic sugar; they have no well-typing of their own.  Well-typing (\cref{def:well-typed-program}) is defined only after expansion (\cref{sec:param-expansion}), on a typed GLP program $(Cs,D)$ (\cref{def:typed-glp-program}) in which no free type parameter remains: every type reached is a concrete user-defined or primitive type.  The unit of type checking is the program: a checker expands it---collecting the instantiation set above and replacing every parameterised reference by its expanded monomorphic name---and applies \cref{def:well-typed-program} to the resulting $(Cs,D)$.  A fragment that still contains a free type parameter after expansion is not a program and is not type-checked.  Checking a parameterised procedure with its parameter left free---equivalently, treating the parameter as the wildcard \verb|_|---is unsound, because a clause that inspects the parameter is then accepted vacuously.  A parameterised procedure acquires concrete types, and is checked, only within a program that instantiates it; with every type then concrete, the variable-pair conditions of \cref{def:well-typed-clause} apply, so a writer/reader mode mismatch in an instantiated clause is rejected.

\ifappendix
\begin{definition}[Program, Linked Program]
\label{def:program}
A \temph{program} is a finite set of modules, scoped as a hierarchy (\cref{sec:modules}), with one or more concrete initial goals.  Its \temph{linked program} is the typed GLP program (\cref{def:typed-glp-program}) obtained by parameterised-type expansion (\cref{sec:param-expansion}), procedure instantiation (\cref{sec:param-procedures}), and linking (\cref{sec:static-linking}), defined only when expansion leaves no free type parameter.
\end{definition}

\begin{definition}[Well-Typed Program]
\label{def:well-typed-program-modular}
A program is \temph{well-typed} if its linked program is well-typed (\cref{def:well-typed-program}).
\end{definition}

Soundness (\cref{thm:soundness}) is a property of the linked program, hence of the program, and makes no claim about a module checked outside a program.
\fi

\subsection{Modular Checking via Abstract Parameters}
\label{sec:abstract-parameters}

The per-instantiation rule above checks a parameterised procedure once for each concrete type its program supplies, so a procedure with no caller in its program goes unchecked.  A procedure that never inspects its parameters can instead be checked once, for all instantiations at once.  Treating a parameter as the wildcard \verb|_| to this end is unsound (\cref{sec:programs-and-modules}); treating it as an \emph{abstract type} is sound, and enforces the parametricity discipline: the procedure passes a parameter-typed value through, and matches no functor or constant against it.

\begin{definition}[Abstract Type]
\label{def:abstract-type}
An \temph{abstract type} is a type-automaton state with no outgoing transitions, distinct from the wildcard \verb|_|.
\end{definition}

By \cref{def:consistent-paths}, a moded term path is consistent with an abstract type at a position exactly when the term there is a variable of the position's mode: a variable is consistent with any same-mode type symbol (rows~2 and~3), whereas a functor or constant has no transition to take and matches no wildcard.

\begin{definition}[Parametrically Well-Typed]
\label{def:parametrically-well-typed}
The \temph{abstract instance} of a parameterised procedure declaration replaces each type parameter by a distinct abstract type.  A parameterised procedure is \temph{parametrically well-typed} if, by its abstract instance, its clauses are well-typed (\cref{def:well-typed-clause}) and every input path is accepted by some clause (\cref{def:input-accepting-clause}).
\end{definition}

\begin{lemma}[Parametricity]
\label{lem:parametricity}
Let a parameterised procedure be parametrically well-typed, with no type parameter occurring as an alternative of a type definition.  Then for every instantiation, by the monomorphic declaration the instantiation produces, the clauses of the instantiated procedure are well-typed (\cref{def:well-typed-clause}) and every input path is accepted by some clause (\cref{def:input-accepting-clause}).
\end{lemma}

\ifappendix
\begin{proof}
Let $\sigma$ map each abstract parameter to its type at the instantiation.  Take a clause well-typed by the abstract instance and a moded-head or body-goal term path of it.  At a position typed by a concrete type, $\sigma$ leaves the type unchanged, so consistency (\cref{def:consistent-paths}) is preserved.  At a position typed by an abstract parameter the term is a variable of the position's mode; $\sigma$ replaces the parameter by a type of the same mode, with which a variable is consistent (rows~2 and~3), so consistency is preserved.  Each variable pair dual by the abstract instance stays dual under $\sigma$, which acts equally on a parameter and its complement.  A type definition's determinism---distinct top-level functors---can change under $\sigma$ only where a parameter is an alternative, excluded by hypothesis, so every type definition stays deterministic.  Input-path acceptance transfers by the same step: a clause accepting an abstract-instance input path carries a variable at each abstract-parameter position on it, and a variable accepts every refinement of that path under $\sigma$.  Hence each instantiated clause is well-typed and every input path of the instantiation is accepted.
\end{proof}
\fi

A parametrically well-typed procedure is thus checked once and certified for every instantiation; the linked program (\cref{sec:programs-and-modules}) need not re-check it, and soundness (\cref{thm:soundness}) follows.  A procedure that inspects a parameter---placing a functor or constant at a parameter position---is not parametrically well-typed, and is checked per instantiation.  A parameter occurring as an alternative of a type definition is likewise checked per instantiation, its determinism resting on the instantiation.  The abstract-parameter route is therefore the sole means of certifying a parametric procedure outside a program: a parameter-inspecting procedure has no well-typing of its own, acquiring one only per instantiation, so a checker presented with it and no instantiation has nothing to certify.

\subsection{Subtyping}
\label{sec:param-subtyping}

Since parameterised types expand to monomorphic types, subtyping is handled by the DFA-based subtype check (Section~\ref{sec:subtyping}). 
\mypara{Example: subtyping through expansion}
Suppose:

\begin{verbatim}
Msg ::= text(String) ; cmd(String).
Cmd ::= cmd(String).
\end{verbatim}

Then \verb|Cmd| $<:$ \verb|Msg| by the standard subtype check.  Expanding \verb|Stream(Cmd)| and \verb|Stream(Msg)| yields:

\begin{verbatim}
Stream<Cmd> ::= [] ; [Cmd | Stream<Cmd>].
Stream<Msg> ::= [] ; [Msg | Stream<Msg>].
\end{verbatim}

The DFA-based subtype check determines that \verb|Stream<Cmd>| $<:$ \verb|Stream<Msg>|, because at each list-cons step, the head type \verb|Cmd| is a subtype of \verb|Msg|.  This is the standard covariance for output positions, with no variance annotation.

\mypara{Contravariance from modes}
For input positions, the mode annotation \verb|?| reverses the subtype direction through complementation.  Consider:

\begin{verbatim}
Channel(In, Out) ::= ch(In, Out?).
\end{verbatim}

A \verb|Channel(Msg, Cmd)| expands to \verb|ch(Msg, Cmd?)|.  A \verb|Channel(Msg, Msg)| expands to \verb|ch(Msg, Msg?)|.  Since \verb|Cmd| $<:$ \verb|Msg|, by complementation \verb|Msg?| $<:$ \verb|Cmd?|.  Therefore \verb|Channel(Msg, Msg)| $<:$ \verb|Channel(Msg, Cmd)|.  This is the standard contravariance, obtained by complementation.

\ifappendix
\subsection{Interaction with Modules}
\label{sec:param-modules}

\mypara{Imported declarations instantiate parameters}
A module that imports \verb|merge| instantiates its type parameter to the specific message type used locally:

\begin{verbatim}
CounterCall ::= inc ; dec ; read(Integer).

imported procedure merge(Stream(CounterCall)?,
                         Stream(CounterCall)?,
                         Stream(CounterCall)).
\end{verbatim}

The \verb|merge| module need not know about \verb|CounterCall|.  Its parameterised declaration \verb|merge(Stream(X)?, Stream(X)?, Stream(X))| is instantiated to \verb|Stream(CounterCall)| at the importing site.

\mypara{Structural compatibility across modules}
Because type identity is structural (Section~\ref{sec:mod-design}), independently expanded types are compatible if their type automata are equivalent or related by subtyping.  If module~A expands \verb|Stream(Msg)| and module~B independently defines \verb|MsgStream ::= [] ; [Msg | MsgStream].|, the two types have identical automata and are compatible.

\mypara{Example: monitor with parameterised merge}
Consider a monitor that accepts requests via a merged input stream:

\begin{verbatim}
%% counter_monitor.glp
CounterCall ::= inc ; dec ; read(Integer).

imported procedure merge(Stream(CounterCall)?,
                         Stream(CounterCall)?,
                         Stream(CounterCall)).

exported procedure counter(Stream(CounterCall)?).
counter(In) :- ...
\end{verbatim}

A client module imports the monitor and provides its own typed stream:

\begin{verbatim}
%% client.glp
CounterCall ::= inc ; dec ; read(Integer).

imported procedure counter(Stream(CounterCall)?).

procedure client(Stream(CounterCall)).
client([inc, inc, read(R?) | Rest?]) :- ...
\end{verbatim}

Both modules independently expand \verb|Stream(CounterCall)| to the same monomorphic type.

\subsection{Expansion Algorithm}
\label{sec:param-algorithm}

The expansion algorithm runs as a preprocessing step, after parsing and before type automaton construction.

\begin{enumerate}
\item \temph{Collect templates.}  Scan all type definitions.  A definition whose name is followed by a parenthesised parameter list is a template; all others are monomorphic.  Templates are recorded but not added to the type environment.

\item \temph{Collect instantiations.}  Scan all type definitions, procedure declarations (including \verb|imported| and \verb|exported| declarations), and type definition bodies for parameterised type references $T(S_1, \ldots, S_k)$.  Each distinct instantiation is recorded.  A type definition body may reference a different parameterised type (e.g., a \verb|Pair| appearing inside \verb|Stream|); such cross-references are collected in this step and expanded in the next.  Scan also every clause body for calls to a parameterised procedure; each call, with the enclosing clause's parameters bound, records an instantiation of the callee, except that a call to a procedure already being instantiated on the current cycle reproduces that instantiation rather than recording a new one.  An instantiated procedure's body may call further parameterised procedures, so repeat the clause-body scan over each new procedure instantiation until none is added; recursion records no instantiation, so this terminates.

\item \temph{Expand.}  For each recorded instantiation, generate a fresh monomorphic type definition by substituting parameters and renaming.  Nested instantiations are expanded inside-out.

\item \temph{Replace references.}  In all type definitions and procedure declarations, replace every parameterised type reference $T(S_1, \ldots, S_k)$ with its expanded name $T\langle S_1, \ldots, S_k \rangle$.

\item \temph{Remove templates.}  Remove parameterised type definitions from the type environment.
\end{enumerate}
\fi

\ifappendix
%==============================================================================
\section{Related Work}
\label{sec:related-work}
%==============================================================================

\mypara{Concurrent Logic Programming}
GLP belongs to the family of concurrent logic programming languages that emerged in the 1980s: Concurrent Prolog~\cite{shapiro1983subset}, GHC~\cite{ueda1986guarded}, and PARLOG~\cite{clark1986parlog}.  These languages interpret goals as concurrent processes communicating through shared logical variables, using committed-choice execution with guarded clauses.

A key evolution was \emph{flattening}: restricting guards to primitive tests only.  Flat Concurrent Prolog (FCP)~\cite{mierowsky1985fcp} and Flat GHC~\cite{ueda1986guarded} demonstrated that flat guards suffice for practical parallel programming while simplifying semantics and implementation.  Shapiro's survey~\cite{shapiro1989family} documents this family and its design space.

GLP can be understood as Flat Concurrent Prolog with the Single-Reader Single-Writer (SRSW) constraint added.  FCP introduced read-only annotations (\verb|?|) distinguishing readers from writers of shared variables, enabling dataflow synchronization.  However, read-only unification proved semantically problematic: Levi and Palamidessi~\cite{levi1985readonly} showed it is order-dependent, and Mierowsky et al.\ documented non-modularity issues.  GHC dispensed with read-only annotations entirely, relying on guard suspension semantics.

Tick~\cite{tick1995deevolution} surveyed this evolution as a ``deevolution'' of concurrent logic programming languages, driven by the growing perception that reduced expressiveness is sufficient for practical programming.  GLP continues this trajectory: the SRSW constraint---requiring that each variable has exactly one writer and one reader occurrence---resolves the difficulties of read-only unification by ensuring that (\ia)~no races occur on variable assignment, and (\ib)~term matching suffices, eschewing unification.  The result is a simpler semantic foundation while preserving the expressiveness of FCP, including streams, hollow messages, and meta-programming.

Janus~\cite{saraswat1990janus} (see also~\cite{debray1993qdjanus}), a concurrent constraint language designed for distributed programming, enforces a two-occurrence restriction on variables---each variable has exactly one writable occurrence (teller) and one readable occurrence (asker)---which is the constraint-programming counterpart of GLP's SRSW restriction.  Lafont's Interaction Nets~\cite{lafont1990interaction} enforce linearity through graph rewriting: each agent has a single principal port, and interaction rules fire only between agents connected at their principal ports, ensuring deterministic reduction.  Both Janus and Interaction Nets share with GLP the design principle that restricting the use of shared resources---whether logic variables, constraint store entries, or graph ports---to single-reader single-writer disciplines yields simpler semantics and enables distributed execution.

\mypara{Modes in Concurrent Logic Programming}
PARLOG used mode declarations at the predicate level, with input modes enforcing one-way matching where clause variables can be instantiated but goal variables cannot be bound.

Ueda's work on moded Flat GHC~\cite{ueda1994moded,ueda1995io} is most directly relevant to Typed GLP.  His mode system assigns polarity to every variable occurrence: positive for input/read capability, negative for output/write capability.  The \emph{well-modedness} property guarantees each variable is written exactly once---a single-writer constraint matching half of GLP's SRSW requirement.  Ueda's subsequent linearity analysis~\cite{ueda2001resource} identifies variables read exactly once, enabling compile-time garbage collection for single-reader structures.

GLP enforces both single-reader and single-writer universally as a syntactic restriction, whereas Ueda's system guarantees single-writer with single-reader as an optional refinement.  This stronger restriction enables GLP's simpler distributed execution model.  Our contribution is a \emph{type system} for this discipline, building on LP types rather than mode inference.  To the best of our knowledge, moded types are the first type system for a concurrent logic language whose communication directionality is part of the type structure, obtained as an abstraction of a moded-atom semantics rather than from a separate mode analysis or inference.

\mypara{Types for Logic Programs}
The type system for GLP builds on the Frühwirth \emph{et al.} framework~\cite{fruhwirth1991lics}, which defines LP types as regular sets of paths and characterises well-typing via path abstraction of the ground-atom semantics.  The companion paper~\cite{yardeni1991jlp} establishes the semantic foundations through tuple-distributive sets.  Our moded types extend this framework by augmenting paths with mode annotations.

Earlier work on LP types includes Mycroft and O'Keefe's prescriptive polymorphic type system~\cite{mycroft1984polymorphic} and Mishra's foundational work on tuple-distributivity~\cite{mishra1984types}.  Mercury~\cite{somogyi1996mercury} provides a practical system separating type checking, mode checking, and determinism inference.  Our work differs in integrating modes directly into the type structure rather than treating them as a separate analysis.

\mypara{Linear Logic, Session Types, and Futures}
GLP's SRSW discipline is reminiscent of linear logic's~\cite{girard1987linear} resource sensitivity: each variable assignment is produced at most once and consumed at most once.  A GLP reader/writer pair is operationally a future or promise~\cite{baker1977future,friedman1976impact}: the writer creates a promise for a value, the reader awaits fulfilment.  Unlike futures in mainstream languages, GLP futures can carry logical terms containing further reader/writer pairs, enabling recursive communication structures---a capability shared with channel-passing calculi such as the $\pi$-calculus~\cite{milner1999communicating}.

Session types~\cite{honda1993types,wadler2012propositions} use linear logic to type communication protocols.  Since GLP messages may contain variables, they allow continuation of the session in both directions, ad infinitum: a reader embedded in a message allows a further message in the same direction, and a writer allows a message in the opposite direction---each of which may contain additional readers and writers.  Thus a \verb|Stream| type is a session: each \verb=[X|Xs]= message consists of an element \verb|X| and continuation \verb|Xs|, extending the interaction indefinitely.  The \verb|CounterCall| example (Section~\ref{sec:typed-glp-examples}) is richer: \verb|read(Integer?)| is a message that embeds a writer that inverts the communication direction, enabling the monitor to respond---typed precisely by the mode inversion in the moded type definition.  GLP's moded types are thus the logic-programming counterpart of session types, with the mathematical foundations differing in derivation: session types from linear logic sequents, moded types from regular sets of moded paths.

Session type theory distinguishes \emph{duality} from \emph{subtyping}.  Duality~\cite{honda1993types} is an involution $S \mapsto \overline{S}$ that swaps input and output modalities.  Subtyping~\cite{gay2005subtyping} is a preorder for safe substitution: $S <: T$ if a channel of type $S$ can be used where $T$ is expected.  Session subtyping is contravariant in input types and covariant in output types.

GLP's moded types incorporate both duality and subtyping directly into the type structure: the $(\cdot)?$ operator relates producer and consumer views of the same channel, analogous to session duality, while subtyping (Section~\ref{sec:subtyping}) captures when interaction is safe despite types not being exact duals.  The well-typing conditions---containment for outputs, coverage for inputs---embody the same variance principles that session subtyping enforces.

Chen and Pfenning's CoLF$_{\omega_1}$~\cite{chen2025colf} interprets logic variables as communication channels, compiling mode-checked programs to a concurrent language.  Their uniqueness requirement enables stream transducers but precludes availability-driven composition such as fair merge.  Systems like SILL$_S$~\cite{balzer2017manifest} extend session types with shared state through explicit acquire/release protocols; GLP achieves sharing through merge processes that combine multiple input streams.

\mypara{Specification-First AI Programming}
The development discipline employed in this work---jointly agreeing on types, procedure declarations, and informal descriptions before AI attempts implementation---synthesises classical specification-first methodologies with recent advances in AI-assisted programming.  Design by Contract~\cite{meyer1997object} prescribes formal interface specifications (preconditions, postconditions, invariants) before implementation.  Literate Programming~\cite{knuth1984literate} advocates natural language exposition alongside code.  Type-driven development~\cite{brady2017type} uses types to guide implementation, with typed holes allowing incremental refinement.  Recent work extends these ideas to LLM-based code generation: Blinn et al.~\cite{blinn2024typed} demonstrate that typed holes provide semantic context for AI code completion, while M\"{u}ndler et al.~\cite{mundler2025type} show that type constraints during generation reduce compilation errors by over 50\%.  Our discipline applies these principles at the specification level: types and procedure declarations serve as a contract between human designer and AI programmer, constraining the space of valid implementations before generation begins.

\fi
%==============================================================================
\section{Conclusion}
\label{sec:conclusion}
%==============================================================================

Our \emph{moded-atom semantics} extends the standard Herbrand atom semantics with modes and with meaning for partial computations, and our \emph{moded types} are regular sets of moded paths over it.  Our central result (\cref{thm:soundness}) establishes the soundness of well-typing---every output path in a well-typed program's well-typed moded-atom semantics conforms to its declared type---and \cref{lem:well-typing-preservation} shows well-typing is preserved through GLP computation.  \cref{prop:input-coverage} gives the contravariant counterpart: every declared input path is matched by some clause head.

\ifappendix We extended the core type system with two mechanisms for programming in the large.  The \emph{module system} (Section~\ref{sec:modules}) supports separate type checking of modules against locally declared interfaces, with structural type compatibility ensuring that independently developed modules can be composed safely.  Static linking flattens a module hierarchy into a single program via procedure renaming; dynamic activation admits a module arriving at runtime as a value, on equality of the hashes of declaration type automata (Section~\ref{sec:dynamic-linking}).  \emph{Parameterised types} (Section~\ref{sec:parameterized-types}) enable generic definitions---\verb|merge| over \verb|Stream(X)| for any \verb|X|---with variance determining how subtyping propagates through type parameters.\fi

When implementing Typed GLP by AI, gaps surfaced at the code level drove revisions upward to the specification and the definitions.  When using Typed GLP by AI, the main benefit was eliminating a class of mode errors that previously manifested only as silent runtime suspensions.

\ifappendix
\mypara{Future Work}
Implementing dynamic activation---the canonical serialisation of type automata underlying its hashes, module export tables, and the \verb|run| and \verb|find_type| primitives---is future work.  Type checking a parametric dynamically-activated procedure at each instantiation requires dynamic type checking, also left to future work.  The interaction between parameterised types and subtyping across module boundaries deserves further study, particularly for higher-order cases where type parameters themselves carry mode structure.
\fi

\subsubsection*{Acknowledgements.}
I thank Thom Frühwirth for discussions and comments on an earlier version of the paper.

% Bibliography
\bibliography{bib}

@article{keidar2026secure,
  author    = {Idit Keidar and Ehud Shapiro},
  title     = {Secure GLP: Executing Digital Social Contracts without Consensus},
  year      = {2026},
  journal      = {In preparation}
}

@book{plotkin1981structural,
  title={A structural approach to operational semantics},
  author={Plotkin, Gordon D},
  year={1981},
  publisher={Aarhus university}
}

@inproceedings{saraswat1990janus,
  author    = {Vijay A. Saraswat and Kenneth M. Kahn and Jacob Levy},
  title     = {Janus: A Step Towards Distributed Constraint Programming},
  booktitle = {Proceedings of the 1990 North American Conference on Logic Programming (NACLP)},
  pages     = {431--446},
  publisher = {MIT Press},
  year      = {1990}
}

@article{debray1993qdjanus,
  author    = {Saumya K. Debray},
  title     = {{QD-Janus}: A Sequential Implementation of {Janus} in {Prolog}},
  journal   = {Software: Practice and Experience},
  volume    = {23},
  number    = {12},
  pages     = {1337--1360},
  year      = {1993}
}

@inproceedings{lafont1990interaction,
  author    = {Yves Lafont},
  title     = {Interaction Nets},
  booktitle = {Conference Record of the 17th {ACM} {SIGPLAN-SIGACT} Symposium on Principles of Programming Languages ({POPL})},
  pages     = {95--108},
  publisher = {ACM},
  year      = {1990}
}

@article{tick1995deevolution,
  author    = {Evan Tick},
  title     = {The Deevolution of Concurrent Logic Programming Languages},
  journal   = {The Journal of Logic Programming},
  volume    = {23},
  number    = {2},
  pages     = {89--123},
  year      = {1995}
}

@article{shapiro2026types,
  title={Types for Grassroots Logic Programs (Full version)},
  author={Shapiro, Ehud},
  journal={arXiv preprint arXiv:2601.17957},
  year={2026}
}

@article{friedman1976impact,
  author = {Friedman, Daniel P. and Wise, David S.},
  title = {The Impact of Applicative Programming on Multiprocessing},
  journal = {Indiana University Computer Science Department Technical Report},
  number = {TR-26},
  year = {1976}
}

@misc{dart2024,
     author       = {{Google}},
     title        = {Dart Programming Language},
     howpublished = {\url{https://dart.dev}},
     year         = {2024}
   }

@book{meyer1997object,
  author    = {Meyer, Bertrand},
  title     = {Object-Oriented Software Construction},
  edition   = {2nd},
  publisher = {Prentice Hall},
  year      = {1997},
  isbn      = {0-13-629155-4},
  note      = {Introduces Design by Contract}
}

@article{knuth1984literate,
  author    = {Knuth, Donald E.},
  title     = {Literate Programming},
  journal   = {The Computer Journal},
  volume    = {27},
  number    = {2},
  pages     = {97--111},
  year      = {1984},
  publisher = {Oxford University Press},
  doi       = {10.1093/comjnl/27.2.97}
}

@book{brady2017type,
  author    = {Brady, Edwin},
  title     = {Type-Driven Development with {I}dris},
  publisher = {Manning Publications},
  year      = {2017},
  isbn      = {978-1617293023}
}

@inproceedings{mundler2025type,
  author    = {M\"{u}ndler, Niels and Guerraoui, Rachid and Vechev, Martin},
  title     = {Type-Constrained Code Generation with Language Models},
  booktitle = {Proceedings of the 46th ACM SIGPLAN Conference on Programming Language Design and Implementation (PLDI)},
  publisher = {ACM},
  year      = {2025},
  note      = {Available at \url{https://arxiv.org/abs/2504.09246}}
}

@inproceedings{blinn2024typed,
  author    = {Blinn, Andrew and Li, Xiang and Kim, June Hyung and Omar, Cyrus},
  title     = {Statically Contextualizing Large Language Models with Typed Holes},
  booktitle = {Proceedings of the ACM on Programming Languages (OOPSLA)},
  volume    = {8},
  number    = {OOPSLA2},
  pages     = {1--29},
  publisher = {ACM},
  year      = {2024},
  doi       = {10.1145/3689746}
}

@article{fowler2025sdd,
  title   = {Understanding Spec-Driven-Development: Kiro, spec-kit, and Tessl},
  author  = {Fowler, Martin},
  journal = {MartinFowler.com},
  year    = {2025},
  month   = {October},
  day     = {15},
  url     = {https://martinfowler.com/articles/exploring-gen-ai/sdd-3-tools.html},
  note    = {Analyzes the shift toward using formal specifications and types as the primary interface for AI code generation}
}

@article{chen2025colf,
  title={CoLF Logic Programming as Infinitary Proof Exploration},
  author={Chen, Zhibo and Pfenning, Frank},
  journal={arXiv preprint arXiv:2510.12302},
  year={2025}
}

@article{mierowsky1985fcp,
  title={On the implementation of Flat Concurrent Prolog},
  author={Mierowsky, C. and Taylor, S. and Shapiro, E. and Levy, J. and Safra, M.},
  journal={Proceedings of the 1985 Symposium on Logic Programming},
  pages={276--286},
  year={1985},
  publisher={IEEE}
}

@article{ueda1994moded,
  doi = {10.1007/BF03038307},
  title={Moded Flat GHC and Its Message-Oriented Implementation Technique},
  author={Ueda, Kazunori and Morita, Masao},
  journal={New Generation Computing},
  volume={12},
  number={4},
  pages={337--368},
  year={1994}
}

@inproceedings{ueda1995io,
  doi = {10.1007/BFb0026579},
  title={I/O mode analysis in concurrent logic programming},
  author={Ueda, Kazunori and Morita, Masao},
  booktitle={Proceedings of the International Symposium on Theory and Practice of Parallel Programming},
  pages={356--368},
  year={1995},
  publisher={Springer}
}

@article{ueda2001resource,
  doi = {10.1007/3-540-45500-0\_5},
  title={Resource-passing concurrent programming},
  author={Ueda, Kazunori},
  journal={Proceedings of TACS 2001},
  pages={95--126},
  year={2001},
  publisher={Springer}
}

@book{pierce2002types,
  author    = {Pierce, Benjamin C.},
  title     = {Types and Programming Languages},
  publisher = {MIT Press},
  year      = {2002},
  isbn      = {0-262-16209-1}
}

@inproceedings{levi1985readonly,
  author = {Levi, Giorgio and Palamidessi, Catuscia},
  title = {The Semantics of the Read-Only Variable},
  booktitle = {Proc. Symposium on Logic Programming},
  publisher = {IEEE},
  pages = {128--137},
  year = {1985}
}

@inproceedings{yardeni1991polymorphically,
  title={Polymorphically Typed Logic Programs},
  author={YARDENI, E and FRUEHWIRTH, T and SHAPIRO, E},
  booktitle={Logic Programming: Proceedings of the Eighth International Conference},
  pages={379--393},
  year={1991},
  organization={MIT Press}
}

@article{cardelli1985understanding,
  author    = {Cardelli, Luca and Wegner, Peter},
  title     = {On Understanding Types, Data Abstraction, and Polymorphism},
  journal   = {Computing Surveys},
  volume    = {17},
  number    = {4},
  pages     = {471--523},
  year      = {1985},
  publisher = {ACM},
  doi       = {10.1145/6041.6042}
}

@article{gay2005subtyping,
  author    = {Gay, Simon and Hole, Malcolm},
  title     = {Subtyping for Session Types in the Pi Calculus},
  journal   = {Acta Informatica},
  volume    = {42},
  number    = {2--3},
  pages     = {191--225},
  year      = {2005},
  publisher = {Springer},
  doi       = {10.1007/s00236-005-0177-z}
}

@article{balzer2017manifest,
  author    = {Balzer, Stephanie and Pfenning, Frank},
  title     = {Manifest Sharing with Session Types},
  journal   = {Proceedings of the ACM on Programming Languages},
  volume    = {1},
  number    = {ICFP},
  articleno = {37},
  pages     = {37:1--37:29},
  year      = {2017},
  publisher = {ACM},
  doi       = {10.1145/3110281}
}

@book{milner1999communicating,
  author    = {Milner, Robin},
  title     = {Communicating and Mobile Systems: The $\pi$-Calculus},
  publisher = {Cambridge University Press},
  year      = {1999},
  isbn      = {0-521-65869-1}
}

@inproceedings{wadler2012propositions,
  author    = {Wadler, Philip},
  title     = {Propositions as Sessions},
  booktitle = {Proceedings of the 17th ACM SIGPLAN International Conference on Functional Programming (ICFP)},
  pages     = {273--286},
  publisher = {ACM},
  year      = {2012},
  doi       = {10.1145/2364527.2364568}
}

@article{yardeni1991jlp,
  author    = {Yardeni, Eyal and Shapiro, Ehud},
  title     = {A Type System for Logic Programs},
  journal   = {Journal of Logic Programming},
  volume    = {10},
  number    = {2},
  pages     = {125--153},
  year      = {1991},
  publisher = {Elsevier},
  doi       = {10.1016/0743-1066(91)90013-G}
}

@inproceedings{fruhwirth1991lics,
  author    = {Fr{\"u}hwirth, Thom and Shapiro, Ehud and Vardi, Moshe Y. and Yardeni, Eyal},
  title     = {Logic Programs as Types for Logic Programs},
  booktitle = {Proceedings of the 6th Annual IEEE Symposium on Logic in Computer Science (LICS)},
  pages     = {300--309},
  year      = {1991},
  publisher = {IEEE Computer Society},
  doi       = {10.1109/LICS.1991.151654}
}

@inproceedings{mishra1984types,
  author    = {Mishra, Prateek},
  title     = {Towards a Theory of Types in Prolog},
  booktitle = {Proceedings of the 1984 International Symposium on Logic Programming},
  pages     = {289--298},
  year      = {1984},
  publisher = {IEEE}
}

@article{somogyi1996mercury,
  author    = {Somogyi, Zoltan and Henderson, Fergus and Conway, Thomas},
  title     = {The Execution Algorithm of Mercury: An Efficient Purely Declarative Logic Programming Language},
  journal   = {Journal of Logic Programming},
  volume    = {29},
  number    = {1--3},
  pages     = {17--64},
  year      = {1996},
  doi       = {10.1016/S0743-1066(96)00068-4}
}

@article{girard1987linear,
  author    = {Girard, Jean-Yves},
  title     = {Linear Logic},
  journal   = {Theoretical Computer Science},
  volume    = {50},
  number    = {1},
  pages     = {1--101},
  year      = {1987},
  publisher = {Elsevier},
  doi       = {10.1016/0304-3975(87)90045-4}
}

@inproceedings{honda1993types,
  author    = {Honda, Kohei},
  title     = {Types for Dyadic Interaction},
  booktitle = {Proceedings of the 4th International Conference on Concurrency Theory (CONCUR)},
  series    = {Lecture Notes in Computer Science},
  volume    = {715},
  pages     = {509--523},
  year      = {1993},
  publisher = {Springer}
}

@article{mycroft1984polymorphic,
  author    = {Mycroft, Alan and O'Keefe, Richard A.},
  title     = {A Polymorphic Type System for {Prolog}},
  journal   = {Artificial Intelligence},
  volume    = {23},
  number    = {3},
  pages     = {295--307},
  year      = {1984},
  publisher = {Elsevier},
  doi       = {10.1016/0004-3702(84)90017-1}
}

@article{shapiro2025glp,
  title={GLP: A Grassroots, Multiagent, Concurrent, Logic Programming Language for AI (full version)},
  author={Shapiro, Ehud},
  journal={arXiv preprint arXiv:2510.15747, Summary to appear in Proc. of ICLP'26},
  year={2025}
}

@inproceedings{ueda1986guarded,
  doi = {10.1007/3-540-16479-0\_17},
  title={Guarded Horn Clauses},
  author={Ueda, Kazunori},
  booktitle={Logic Programming '85},
  series={Lecture Notes in Computer Science},
  volume={221},
  pages={168--179},
  year={1986},
  publisher={Springer}
}

@article{robinson1965machine,
  title={A machine-oriented logic based on the resolution principle},
  author={Robinson, John Alan},
  journal={Journal of the ACM (JACM)},
  volume={12},
  number={1},
  pages={23--41},
  year={1965},
  publisher={ACM New York, NY, USA}
}

@inproceedings{shapiro2025atomic,
  title={Grassroots Platforms with Atomic Transactions: Social Graphs, Cryptocurrencies, and Democratic Federations},
  author={Shapiro, Ehud},
  booktitle={Proceedings of the 27th International Conference on Distributed Computing and Networking},
  pages={71--81},
  doi = {10.1145/3772290.3772309},
  note = {arXiv preprint arXiv:2502.11299},
  year={2026}
}

@article{shapiro1984alternation,
  title={Alternation and the computational complexity of logic programs},
  author={Shapiro, Ehud},
  journal={The Journal of Logic Programming},
  volume={1},
  number={1},
  pages={19--33},
  year={1984},
  publisher={Elsevier}
}

@inproceedings{gaifman1989fully,
  title={Fully abstract compositional semantics for logic programs},
  author={Gaifman, Haim and Shapiro, Ehud},
  booktitle={Proceedings of the 16th ACM SIGPLAN-SIGACT symposium on Principles of programming languages},
  pages={134--142},
  year={1989}
}

@article{clark1986parlog,
  doi = {10.1145/5001.5390},
  title={PARLOG: parallel programming in logic},
  author={Clark, Keith and Gregory, Steve},
  journal={ACM Transactions on Programming Languages and Systems (TOPLAS)},
  volume={8},
  number={1},
  pages={1--49},
  year={1986},
  publisher={ACM New York, NY, USA}
}

@article{shapiro1983subset,
  title={A subset of Concurrent Prolog and its interpreter},
  author={Shapiro, Ehud},
  journal={ICOT Technical Report, TR-003},
  year={1983}
}

@inproceedings{baker1977future,
  author    = {Henry G. Baker and Carl Hewitt},
  title     = {The Incremental Garbage Collection of Processes},
  booktitle = {Proceedings of the 1977 Symposium on Artificial Intelligence and Programming Languages},
  year      = {1977},
  pages     = {55--59},
  publisher = {ACM},
  doi       = {10.1145/800228.806932}
}

@inproceedings{shapiro2023grassrootsBA, 
author={Shapiro, Ehud},
 title={Grassroots Distributed Systems: Concept, Examples, Implementation and Applications (Brief Announcement)
},year = {2023},
publisher = {LIPICS},
booktitle = {37th International Symposium on Distributed Computing (DISC 2023). (Extended version: arXiv:2301.04391)},
address = {Italy},
notes={Extended version: arXiv preprint arXiv:2301.04391},
pages = {47:1, 47:7}
}

@article{shapiro1989family,
  doi = {10.1145/72551.72555},
  title={The family of concurrent logic programming languages},
  author={Shapiro, Ehud},
  journal={ACM Computing Surveys (CSUR)},
  volume={21},
  number={3},
  pages={413--510},
  year={1989},
  publisher={ACM New York, NY, USA}
}

@misc{proof,
  title={Proof of Stake FAQ, Ethereum Wiki},
  author={Nakamoto, Satoshi and Bitcoin, A},
  howpublished = { \\ \MYhref{https://eth.wiki/en/concepts/proof-of-stake-faqs}{https://eth.wiki/en/concepts/proof-of-stake-faqs}},
  year={2019}
}

@incollection{silverman1988logix,
  title={The Logix system user manual Version 1.21},
  author={Silverman, William and Hirsch, Michael and Houri, Avshalom and Shapiro, Ehud},
  booktitle={Concurrent Prolog: Collected Papers},
  pages={46--77},
  year={1988}
}

% Appendices (arXiv build only; LOPSTR build omits them)
\ifappendix
\appendix

\section{Deferred Proofs}
\label{app:proofs}

This appendix contains proofs of propositions from Sections~\ref{sec:lp} and~\ref{sec:glp}.

\propLPDeduction*

\begin{proof}
By induction on the length $n$ of the run.

\emph{Base case} ($n=0$): The outcome is $(G_0 \mathrel{\mbox{\texttt{:-}}} G_0)\emptyset = (G_0 \mathrel{\mbox{\texttt{:-}}} G_0)$, which is the tautology $G_0 \Leftarrow G_0$.

\emph{Inductive step}: Suppose the proposition holds for runs of length $n-1$.  Consider a run of length $n$ with final transition $(G_{n-1}, \sigma_{n-1}) \rightarrow (G_n, \sigma_n)$.  By definition, there exist atom $A \in G_{n-1}$ and clause $C = (H \mathrel{\mbox{\texttt{:-}}} B) \in P$ (renamed apart) such that $A$ and $H$ have mgu $\hat\sigma$, $G_n = (G_{n-1} \setminus \{A\} \cup B)\hat\sigma$, and $\sigma_n = \sigma_{n-1} \circ \hat\sigma$.

The clause $C$, being universally quantified, entails the instance $(H \mathrel{\mbox{\texttt{:-}}} B)\hat\sigma$, i.e., $H\hat\sigma \Leftarrow B\hat\sigma$.  Since $A\hat\sigma = H\hat\sigma$ (by unification), we have $A\hat\sigma \Leftarrow B\hat\sigma$.

By the inductive hypothesis, the prefix run has outcome $(G_0 \mathrel{\mbox{\texttt{:-}}} G_{n-1})\sigma_{n-1}$, a logical consequence of $P$.  Applying $\hat\sigma$ and substituting $A\hat\sigma$ with $B\hat\sigma$ (justified by the clause instance), we obtain $(G_0 \mathrel{\mbox{\texttt{:-}}} G_n)\sigma_n$, also a logical consequence of $P$.
\end{proof}

\propGLPDeduction*

\begin{proof}
The $/?$ operator replaces every reader $X?$ by its paired writer $X$, transforming GLP terms into LP terms.  We show that the cGLP run $\rho$ corresponds to an LP run $\rho/?$ of $LP(P/?)$.

Consider a cGLP transition $(G, \sigma) \rightarrow (G', \sigma')$:
\begin{itemize}
    \item \emph{Reduce transition}: Goal $A$ reduces with clause $C$ via writer mgu $\hat\sigma$.  Applying $/?$, the clause $C/?$ is an LP clause, and $A/?$ unifies with the head $H/?$ via the mgu $\hat\sigma/?$ (since writers map to writers).  This is a valid LP reduction.
    \item \emph{Communicate transition}: A reader $X? \in G$ is replaced by the value $T$ assigned to its paired writer.  Under $/?$, both $X?$ and $X$ map to $X$, so this transition becomes the identity---the variable $X$ is already assigned $T$ in the LP view.
\end{itemize}

Thus each cGLP transition corresponds to zero or one LP transitions, and the cGLP run $\rho$ projects to an LP run $\rho/?$ of $LP(P/?)$.  By Proposition~\ref{prop:lp-deduction}, the outcome of $\rho/?$ is a logical consequence of $P/?$.
\end{proof}

\propSOPreservation*

\begin{proof}
By induction on the length of the run.  The base case is immediate: $G_0$ satisfies SO by assumption.

For the inductive step, assume $G$ satisfies SO and consider a transition $(G, \sigma) \rightarrow (G', \sigma')$:
\begin{itemize}
    \item \emph{Reduce transition}: Goal $A \in G$ reduces with clause $C = (H \mathrel{\mbox{\texttt{:-}}} B)$ via writer mgu $\hat\sigma$, yielding $G' = (G \setminus \{A\} \cup B)\hat\sigma$.  Since $C$ satisfies SRSW, it satisfies SO.  Since $C$ is renamed apart from $G$, the variables in $B$ are fresh.  The writer mgu $\hat\sigma$ maps writers in $A$ to subterms of $H$ and vice versa; by SO of both $G$ and $C$, each variable is assigned at most once.  Applying $\hat\sigma$ to $(G \setminus \{A\} \cup B)$ replaces each variable by a term containing fresh variables (from $B$) or ground subterms.  Since no variable in $G \setminus \{A\}$ occurs in $A$ (by SO of $G$), and no variable in $B$ occurs in $G$ (by renaming apart), $G'$ satisfies SO.
    
    \item \emph{Communicate transition}: $G' = G\hat\sigma?$ where $\hat\sigma? = \{X? := T\}$.  Since $G$ satisfies SO, $X?$ occurs at most once in $G$.  Replacing this single occurrence by $T$ (which satisfies SO by Definition~\ref{def:writers-assignment}) preserves SO, provided $T$ shares no variables with the rest of $G$.  By the proper run condition (Definition~\ref{def:cglp-run-outcome}), variables in $T$ are fresh, so $G'$ satisfies SO.
\end{itemize}
\end{proof}

\propMonotonicity*

\begin{proof}
Suppose goal $A$ can reduce with clause $C$ at step $i$, meaning the writer mgu of $A$ and the head $H$ of (a renaming of) $C$ succeeds.  Consider what can change between steps $i$ and $j > i$:
\begin{itemize}
    \item \emph{Reduce transitions on other goals}: These do not affect $A$ directly.  By SO, no other goal shares a writer with $A$, so no other reduction can assign a writer in $A$.
    
    \item \emph{Communicate transitions}: These instantiate readers, not writers.  A communicate transition $X? := T$ may instantiate a reader $X? \in A$, yielding $A' = A\{X? := T\}$.  We show $A'$ can still reduce with $C$:
    
    The original writer mgu succeeded, meaning at position $p$ where $X?$ occurred in $A$, either (a) $H$ had a writer $Y$ at position $p$, yielding assignment $Y := X?$, or (b) $H$ had a reader $Y?$ at position $p$, which would have caused failure (reader-reader), contradiction.
    
    In case (a), after the communicate transition, $A'$ has $T$ at position $p$.  The clause $C$ (renamed apart from $A'$) has a fresh writer $Y'$ at position $p$.  The writer mgu now yields $Y' := T$, which succeeds.
    
    \item \emph{Reduce transition on $A$}: If $A$ itself is reduced at some step $k$ with $i < k \le j$, then an instance of $A$ has been reduced, satisfying the proposition.
\end{itemize}

Thus, if $A$ has not been reduced by step $j$, the (possibly instantiated) goal $A'$ at step $j$ can still reduce with a fresh renaming of $C$.
\end{proof}

\section{GLP Language Primitives and the Root \texorpdfstring{\texttt{self.glp}}{self.glp}}
\label{sec:root-self}

The scope chain of every module (Section~\ref{sec:modules}) starts from the GLP language primitives, followed by the \verb|self.glp| of the hierarchy root.  This section describes both, and the type alias mechanism.

\subsection{GLP Language Primitives}
\label{app:machine-primitives}

The GLP language primitives are the primitive types (\verb|Integer|, \verb|Real|, \verb|String|, \verb|Module|, and \verb|Exp|), guard predicates, body kernels, and system predicates.  They are implemented by the runtime and form the base of every module's scope chain; their complete listing appears in the companion paper~\cite{shapiro2025glp}.  \verb|Number| and \verb|Constant| are not primitive: they are defined in the root \verb|self.glp| below.  Every scope above them---the root \verb|self.glp| and every application module---is expressed in GLP, introduces no further primitive, and reaches runtime functionality only through the provided primitives.

\subsection{Admission to the Primitive Layer}
\label{app:system-mode}

A \emph{reserved constant} is a constant that names a language primitive---a kernel predicate or a reserved functor; such primitives carry a \verb|_| prefix by convention.  It may be named only by a module that declares the directive \verb|-mode(system)|.  A constant that names no primitive is not reserved, even when written with the prefix.  That directive is itself confined to the primitive layer: only the root \verb|self.glp| and the modules under \verb|programs/system/| declare it, and no other module names a reserved constant.  An application module---one implementing a platform, protocol, or program---declares neither, and reaches runtime functionality by calling the system predicates that \verb|programs/system/| exports, as it would any cross-module procedure.  The loader enforces this at load time, not by convention: it rejects a module outside the root \verb|self.glp| and \verb|programs/system/| that declares \verb|-mode(system)|, and a module that names a reserved constant without \verb|-mode(system)|.

\subsection{The Root \texorpdfstring{\texttt{self.glp}}{self.glp}}
\label{app:root-self-glp}

The hierarchy's root \verb|self.glp| is the outermost ancestor scope, shadowable like any other, and everything it provides is expressed in GLP.  It defines four parameterised type templates:

\begin{verbatim}
Stream(X) ::= [] ; [X | Stream(X)].
OpenStream(X) ::= [X | Stream(X)].
DiffList(X) ::= Stream(X) \ Stream(X)?.
Channel(In, Out) ::= ch(In, Out?).
\end{verbatim}

\verb|DiffList(X)| is a difference list, for constant-time concatenation.  \verb|Channel(In, Out)| is a bidirectional communication channel with read stream \verb|In| and write stream \verb|Out?| (the mode annotation on \verb|Out| indicates it is consumed by the channel holder).

Templates are expanded at each use site (Section~\ref{sec:param-expansion}).

It also defines the derived types \verb|Number| and \verb|Constant|:

\begin{verbatim}
Number   ::= Integer ; Real.
Constant ::= Number ; String ; Module.
\end{verbatim}

The empty list \verb|[]| is a \verb|String|, hence a \verb|Constant|.

The root \verb|self.glp| also defines single-unit-clause procedures that serve as defined guards.  When called in guard position, the partial evaluator unfolds them at compile time, before type checking.  The call \verb|X = T| assigns \verb|T| to the writer \verb|X|.

\begin{verbatim}
procedure =(_?, _).
X? = X.

procedure dl_append(DiffList(X)?, DiffList(X)?,
                    DiffList(X)).
dl_append(A\B?, B\C?, A?\C).

procedure dl_to_list(DiffList(X)?, Stream(X)).
dl_to_list(L\[], L?).

procedure new_channel(Channel(X, Y), Channel(Y, X)).
new_channel(ch(Xs?, Ys), ch(Ys?, Xs)).

procedure send(X?, Channel(Y, Stream(X))?,
               Channel(Y, Stream(X))).
send(X, ch(In, [X?|Out?]), ch(In?, Out)).

procedure receive(X, Channel(OpenStream(X), Y)?,
                  Channel(Stream(X), Y)).
receive(X?, ch([X|In], Out?), ch(In?, Out)).

Closed ::= [].

procedure close(Channel(Closed, Closed)?).
close(ch([], [])).
\end{verbatim}

\verb|receive| consumes a message from a non-empty read stream; its continuation's read stream may be empty.  \verb|close| consumes a closed read stream and closes the write stream.  A channel consumer covers its read-stream type with two guarded clauses, one calling \verb|receive| and one \verb|close|: unfolding places the corresponding channel structure in each head, and together the heads accept every input path of the channel (\cref{def:well-typed-program}).

It also defines the parameterised \verb|merge| procedure (Section~\ref{sec:parameterized-types}):

\begin{verbatim}
procedure merge(Stream(X)?, Stream(X)?, Stream(X)).
merge([X|Xs], Ys, [X?|Zs?]) :- merge(Ys?, Xs?, Zs).
merge(Xs, [Y|Ys], [Y?|Zs?]) :- merge(Xs?, Ys?, Zs).
merge(Xs, [], Xs?).
merge([], Ys, Ys?).
\end{verbatim}

\subsection{Type Aliases}
\label{app:type-aliases}

For readability, programs may define type aliases.  A \emph{simple alias} has a single alternative that is a type reference:
\begin{verbatim}
AgentId ::= Constant.       % alias for a defined type
MyStream ::= Stream.        % alias for defined type
ConsumedStream ::= Stream?. % alias for dual of defined type
\end{verbatim}
A \emph{union alias} has multiple alternatives, each of which is a type reference:
\begin{verbatim}
Msg ::= NetMsg ; UserMsg.   % union of two message types
\end{verbatim}
Union aliases are expanded by collecting all alternatives from the referenced types.  For example, if \verb|NetMsg ::= msg(_, _, _).| and \verb|UserMsg ::= sent(_, _) ; received(_, _).|, then \verb|Msg| is expanded to \verb|Msg ::= msg(_, _, _) ; sent(_, _) ; received(_, _).|  The referenced types must be defined (not aliases themselves or primitives), and the expanded alternatives must satisfy the determinism requirement.

During preprocessing, every occurrence of an alias is replaced by its expansion, before type automaton construction.  Circular alias chains (e.g., \verb|A ::= B. B ::= A.|) are rejected.

\section{Type Automaton Construction}
\label{app:type-automaton}

This appendix provides the formal construction of the type automaton summarised in Section~\ref{sec:type-automaton}.

\begin{definition}[Type Automaton]
\label{def:type-automaton}
Given a typed GLP program $P = (Cs, D)$, the \temph{type automaton} $\mathcal{A}_D = (Q, \Sigma, \delta, F)$ is defined as follows:

\mypara{States $Q$}  The state set comprises: (1)~user-defined type states $T$ and $T?$ for each type name; (2)~procedure states $p/n$; (3)~primitive type states (see Appendix~\ref{sec:root-self}); (4)~a final state $\checkmark$.

\mypara{Alphabet $\Sigma$}  Transition labels are tuples $(f, n, i, m)$: functor $f$, arity $n$, argument position $i \in \{1,\ldots,n\}$, and mode $m \in \{\uparrow, \downarrow\}$.  For constants, the label is $(c, 0, m)$.

\mypara{Initial and final states}  The initial state for type $T$ is $T$; for procedure $p/n$, it is $p/n$.  Final states are \verb|_|, \verb|_?|, and $\checkmark$.
\end{definition}

\begin{definition}[Transition Function]
\label{def:transition-function}
The transition function $\delta: Q \times \Sigma \to Q$ is defined by the following rules:

\mypara{Procedure transitions}  For a procedure declaration \verb|p(|$T_1, \ldots, T_n$\verb|)|:
\[
\delta(p/n, (p, n, i, m_i)) = 
\begin{cases}
T_i? & \text{if the declaration writes } T_i? \\
T_i & \text{otherwise}
\end{cases}
\]
The mode $m_i$ is $\downarrow$ for input arguments, $\uparrow$ for output arguments.

\mypara{Type definition transitions}  For a type definition $T ::= A_1 \mathbin{;} \cdots \mathbin{;} A_k$, each alternative $A_j$ contributes transitions from state $T$:
\begin{itemize}
    \item If $A_j$ is a constant $c$: $\delta(T, (c, 0, \uparrow)) = \checkmark$
    \item If $A_j$ is a compound term $f(S_1, \ldots, S_n)$: for each $i \in \{1,\ldots,n\}$,
    \[
    \delta(T, (f, n, i, m_i)) = S_i^{m_i}
    \]
    where $S_i^{m_i}$ denotes $S_i$ if $m_i = \uparrow$, or $S_i?$ if $m_i = \downarrow$ (mode inherited from enclosing context, possibly complemented by embedded $?$).
    \item If $A_j$ references another type $S$ or $S?$: transitions are inherited from $S$ or $S?$ respectively.
\end{itemize}

\mypara{Primitive type transitions}  
\begin{itemize}
    \item $\delta(\verb|Integer|, (k, 0, \uparrow)) = \checkmark$ for any integer literal $k$
    \item $\delta(\verb|Real|, (r, 0, \uparrow)) = \checkmark$ for any real literal $r$
    \item $\delta(\verb|String|, (s, 0, \uparrow)) = \checkmark$ for any string literal $s$
    \item $\delta(\verb|Module|, (\mu, 0, \uparrow)) = \checkmark$ for any module term $\mu$
\end{itemize}
Dual states have the same transitions with mode $\downarrow$.

\mypara{Wildcard states}  \verb|_| accepts any produced term (a writer or ground term with mode $\uparrow$); \verb|_?| accepts any consumed term (a reader or ground term with mode $\downarrow$).  Both are final states, with no outgoing transitions.
\end{definition}

\begin{definition}[Dual Type Automaton]
\label{def:dual-type-automaton}
The automaton for the dual type $T?$ has, for each transition $\delta(T, (f, n, i, m)) = S$, the transition $\delta(T?, (f, n, i, \overline{m})) = S?$, where $\overline{\uparrow} = \downarrow$ and $\overline{\downarrow} = \uparrow$.  Complementation is an involution: $(T?)? = T$.
\end{definition}

\begin{example}[Type Automaton for Stream]
\label{ex:stream-automaton}
For the type definition \verb+Stream ::= [] ; [_|Stream]+, the type automaton has:

\medskip
\noindent\textbf{State \texttt{Stream}} (producer view) with transitions:
\begin{itemize}
    \item $\delta(\verb|Stream|, (\verb|[]|, 0, \uparrow)) = \checkmark$
    \item $\delta(\verb|Stream|, (\verb|"."|, 2, 1, \uparrow)) = \verb|_|$
    \item $\delta(\verb|Stream|, (\verb|"."|, 2, 2, \uparrow)) = \verb|Stream|$
\end{itemize}

\noindent\textbf{State \texttt{Stream?}} (consumer view, dual) with transitions:
\begin{itemize}
    \item $\delta(\verb|Stream?|, (\verb|[]|, 0, \downarrow)) = \checkmark$
    \item $\delta(\verb|Stream?|, (\verb|"."|, 2, 1, \downarrow)) = \verb|_?|$
    \item $\delta(\verb|Stream?|, (\verb|"."|, 2, 2, \downarrow)) = \verb|Stream?|$
\end{itemize}

\end{example}

\begin{example}[Type Automaton for Procedure]
\label{ex:procedure-automaton}
For the declaration \verb|procedure merge(Stream?,Stream?,Stream)|, the automaton includes:

\noindent\textbf{State \texttt{merge/3}} with transitions:
\begin{itemize}
    \item $\delta(\verb|merge/3|, (\verb|merge|, 3, 1, \downarrow)) = \verb|Stream?|$
    \item $\delta(\verb|merge/3|, (\verb|merge|, 3, 2, \downarrow)) = \verb|Stream?|$
    \item $\delta(\verb|merge/3|, (\verb|merge|, 3, 3, \uparrow)) = \verb|Stream|$
\end{itemize}
\end{example}

\mypara{Path acceptance}
A moded path $\pi = (m_0, f_0) \cdot (i_1, m_1, f_1) \cdots (i_k, m_k, f_k)$ determines the label sequence $(f_0, n_0, i_1, m_1) \cdots (f_{k-1}, n_{k-1}, i_k, m_k)$, where $n_j$ is the arity of $f_j$.  The automaton \emph{accepts} $\pi$ from state $q$ if this sequence leads from $q$ to a state $S$ such that either $S = f_k$, or $f_k$ is a constant and $\delta(S, (f_k, 0, m_k)) = \checkmark$.  The set of paths accepted from state $T$ is $paths(T)$.

\fi
\end{document}